\documentclass[11pt]{article}

\usepackage{amssymb}

\usepackage{graphicx}
\usepackage{amsmath}
\usepackage{makeidx}
\usepackage{graphicx}
\usepackage{graphicx}
\usepackage{indentfirst}

\newcounter{resultnum}[section]

\newcounter{conclusionnum}[section]

\newcounter{conditionnum}[section]

\newcounter{conjecturenum}[section]

\newcounter{examplenum}[section]

\newcounter{exercisenum}[section]

\newcounter{lemmanum}[section]

\newcounter{notationnum}[section]

\newcounter{theoremnum}[section]

\newcounter{definitionnum}[section]

\newcounter{corollarynum}[section]

\newcounter{remarknum}[section]

\newcounter{propositionnum}[section]

\newcounter{acknowledgementnum}[section]

\newcounter{algorithmnum}[section]

\newcounter{axiomnum}[section]

\newcounter{casenum}[section]

\newcounter{claimnum}[section]

\newcounter{summarynum}[section]

\newcounter{problemnum}[section]

\begin{document}

\title{Nonholonomic Ricci Flows and Parametric Deformations of the Solitonic pp--Waves and Schwarzschild Solutions}
\date{November 10, 2008}

\author{ Sergiu I. Vacaru
\thanks{
the affiliation for Fields Institute is for a former visiting position;  \newline
Sergiu.Vacaru@gmail.com;\   http://www.scribd.com/people/view/1455460-sergiu
 } \\
{\quad} \\
{\small {\textsl{Faculty of Mathematics, University "Al. I. Cuza" Ia\c si},}
}\\
{\small {\textsl{\ 700506, Ia\c si, Romania}} }\\
{\small and}\\
{\small \textsl{The Fields Institute for Research in Mathematical Science}}
\\
{\small \textsl{222 College Street, 2d Floor, Toronto \ M5T 3J1, Canada}} }

\maketitle

\begin{abstract}
We study Ricci flows of some classes of physically valuable solutions in
Einstein and string gravity. The anholonomic frame method is applied for
generic off--diagonal metric ansatz when the field/ evolution equations are
transformed into exactly integrable systems of partial differential
equations. The integral varieties of such solutions, in four and five
dimensional gravity, depend on arbitrary generation and integration
functions of one, two and/ or three variables. Certain classes of
nonholonomic frame constraints allow us to select vacuum and/or Einstein
metrics, to generalize such solutions for nontrivial string (for instance,
with antisymmetric torsion fields) and matter field sources. A very
important property of this approach (originating from Finsler and Lagrange
geometry but re--defined for semi--Riemannian spaces) is that new classes of
exact solutions can be generated by nonholonomic deformations depending on
parameters associated to some generalized Geroch transforms and Ricci flow
evolution. In this paper, we apply the method to construct in explicit form
some classes of exact solutions for multi--parameter Einstein spaces and
their nonholonomic Ricci flows describing evolutions/interactions of
solitonic pp--waves and deformations of the Schwarzschild metric. We explore
possible physical consequences and speculate on their importance in modern
gravity.

\vskip0.3cm

\textbf{Keywords:}\ Ricci flows, exact solutions, nonholonomic frames,
gra\-vitational solitons, pp--waves, methods of Fin\-s\-ler and Lagrange
geometry, nonlinear connections

\vskip3pt

PACS Classification:

04.20.Jb, 04.30.Nk, 04.50.+h, 04.90.+e, 02.30.Jr, 02.40.-k

\vskip2pt

2000 AMS Subject Classification:

53A99, 53B40, 53C12, 53C44, 83C15, 83C20, 83C99, 83E99
\end{abstract}


\section{Introduction}

This is the fifths paper in a series of works on nonholonomic Ricci flows of
metrics and geometric objects subjected to nonintegrable (nonholonomic)
constraints \cite{5nhrf01,5nhrf02,5nhrf03,5nhrf04}. It is devoted to
explicit applications of new geometric methods in constructing exact
solutions in gravity and Ricci flow theory. Specifically, we shall consider
a set of particular solutions with solitonic pp--waves and anholonomic
deformations of the Schwarzschild metric, defined by generic off--diagonal
metrics\footnote{%
which can not diagonalized by coordinate transforms}, describing nonlinear
gravitational interactions and gravitational models with effective
cosmological constants, for instance, induced by string corrections \cite%
{5string1,5string2} or effective approximations for matter fields. We shall
analyze how such gravitational configurations (some of them generated as
exact solutions by geometric methods in Section 4 of Ref. \cite{5vnhes}) may
evolve under Ricci flows; on previous results and the so--called anholonomic
frame method of constructing exact solutions, see works \cite%
{5vsgg,5valg,5vesnc,5vncg,5vs1,5vs2} and references therein.

Some of the most interesting directions in modern mathematics are related to
the Ricci flow theory \cite{5ham2,5per1}, see reviews \cite%
{5caozhu,5kleiner,5rbook}. There were elaborated a set of applications with
such geometric flows, following low dimensional or approximative methods to
construct solutions of evolution equations, in modern gravity and
mathematical physics, for instance, for low dimensional systems and gravity %
\cite{5ni,5cv,5gk,5osw} and black holes and cosmology \cite{5hw,5bop}. One
of the most important tasks in such directions is to formulate certain
general methods of constructing exact solutions for gravitational systems
under Ricci flow evolution of fundamental geometric objects.

In Refs. \cite{5vrf,5vv1,5vv2}, considering the Ricci flow evolution
parameter as a time like, or extra dimension, coordinate, we provided the
first examples when physically valuable Ricci flow solutions were
constructed following the anholonomic frame method. A quite different scheme
was considered in Ref. \cite{5nhrf03} with detailed proofs that the
information on general metrics and connection in Riemann--Cartan geometry,
and various generalizations to nonholonomic Lagrange--Finsler spaces, can be
encoded into bi--Hamilton structure and related solitonic hierarchies. There
were formulated certain conditions when nonholonomic solitonic equations can
be constrained to extract exact solutions for the Einstein equations and
Ricci flow evolution equations.

The previous partner paper \cite{5nhrf04} was devoted to the geometry of
paramet\-rized nonholonomic frame transforms as superpositions of the Geroch
transforms (generating exact vacuum gravitational solutions with Killing
symmetries, see Refs. \cite{5geroch1,5geroch2}) and the anholonomic frame
deformations and oriented to carry out a program of generating off--diagonal
exact solutions in gravity \cite{5vnhes} and Ricci flow theories. The goal
of this work is to show how such new classes of parametric nonholonomic
solutions, formally constructed for the Einstein and string gravity \cite%
{5vnhes}, can be generalized to satisfy certain geometric evolution
equations and define Ricci flows of physically valuable metrics and
connections.

The structure of the paper is the following:\ In section 2, we outline the
necessary formulas for nonholonomic Einstein spaces and Ricci flows. There
are introduced the general ansatz for generic off--diagonal metrics (for
which, we shall construct evolution/ field exact solutions) and the primary
metrics used for parametric nonholonomic deformations to new classes of
solutions.

In section 3, we construct Ricci flow solutions of solitoninc pp--waves in
vacuum Einstein and string gravity.

Section 4 is devoted to a study of parametric nonholonomic tranforms
(defined as superpositions of the parametric transforms and nonholonomic
frame deformations) in order to generate (multi-) parametric solitonic
pp--waves for Ricci flows and in Einstein spaces.

Section 5 generalizes to Ricci flow configurations the exact solutions
generated by parametric nonholonomic frame transforms of the Schwarz\-schild
metric. There are analyzed deformations and flows of stationary backgrounds,
considered anisotropic polarizations on extra dimension coordinate (possibly
induced by Ricci flows and extra dimension interactions) and examined five
dimensional solutions with running of parameters on nonholonomic time
coordinate and flow parameter.

The paper concludes with a discussion of results in section 6.

The Appendix contains some necessary formulas on effective cosmological
constants and nonholonomic configurations induced from string gravity.

\section{Preliminaries}

We work on five and/or four dimensional, (5D and/ or 4D), nonholonomic
Riemannian manifolds $\mathbf{V}$ of necessary smooth class and conventional
splitting of dimensions $\dim \mathbf{V}=$ $n+m$ for $n=3,$ or $n=2$ and $%
m=2,$ defined by a nonlinear connection (N--connection) structure $\mathbf{N}%
=\{N_{i}^{a}\},$ such manifolds are also called N--anholonomic \cite{5vsgg}.
The local coordinates are labelled in the form $u^{\alpha
}=(x^{i},y^{a})=(x^{1},x^{\widehat{i}},y^{4}=v,y^{5}),$ for $i=1,2,3$ and $%
\widehat{i}=2,3$ and $a,b,...=4,5.$ Any coordinates from a set $u^{\alpha }$
can be for a three dimensional (3D) space, time, or extra dimension (5th
coordinate). Ricci flows of geometric objects will be parametrized by a real
$\chi \in \lbrack 0,\chi _{0}].$ Four dimensional (4D) spaces, when the
local coordinates are labelled in the form $u^{\widehat{\alpha }}=(x^{%
\widehat{i}},y^{a}),$ i. e. without coordinate $x^{1},$ are defined as a
trivial embedding into 5D ones. In general, we shall follow the conventions
and methods stated in Refs. \cite{5nhrf04,5vnhes} (the reader is recommended
to consult those works on main definitions, denotations and geometric
constructions).

\subsection{Ansatz for the Einstein and Ricci flow equations}

A nonholonomic manifold $\mathbf{V},$ provided with a N--connection
(equivalently, with a locally fibred) structure and a related preferred
system of reference, can be described in equivalent forms by two different
linear connections, the Levi Civita $\nabla $ and the canonical
distinguished connection, d--connection $\widehat{\mathbf{D}},$ both
completely defined by the same metric structure
\begin{eqnarray}
\mathbf{g} &=&\mathbf{g}_{\alpha \beta }\left( u\right) \mathbf{e}^{\alpha
}\otimes \mathbf{e}^{\beta }=g_{ij}\left( u\right) e^{i}\otimes
e^{j}+h_{ab}\left( u\right) \ \mathbf{e}^{a}\otimes \ \mathbf{e}^{b},
\label{5dmetr} \\
e^{i} &=&dx^{i},~\mathbf{e}^{a}=dy^{a}+N_{i}^{a}(u)dx^{i},  \notag
\end{eqnarray}%
in metric compatible forms, $\nabla \mathbf{g}=0$ and $\widehat{\mathbf{D}}%
\mathbf{g}=0.$ It should be noted that, in general, $\widehat{\mathbf{D}}$
contains some nontrivial torsion coefficients induced by $N_{i}^{a}$ , see
details in Refs. \cite%
{5nhrf01,5nhrf02,5nhrf03,5nhrf04,5vsgg,5valg,5vesnc,5vncg,5vs1,5vs2}. For
simplicity, we shall omit ''hats'' and indices, or other labels, and write,
for instance, $u=(x,y),$ $\partial _{i}=\partial /\partial x^{i}$ and $%
\partial _{a}=\partial /\partial y^{a},...$ if such simplifications will not
result in ambiguities.

In order to consider Ricci flows of geometric objects, we shall work with
families of ansatz $~^{\chi }\mathbf{g}=\mathbf{g}(\chi ),$ of type (\ref%
{5dmetr}), parametrized by $\chi ,$
\begin{eqnarray}
~^{\chi }\mathbf{g} &=&g_{1}{dx^{1}}\otimes {dx^{1}}+g_{2}(x^{2},x^{3},\chi )%
{dx^{2}}\otimes {dx^{2}}+g_{3}\left( x^{2},x^{3},\chi \right) {dx^{3}}%
\otimes {dx^{3}}  \notag \\
&&+h_{4}\left( x^{k},v,\chi \right) \ ~^{\chi }{\delta v}\otimes ~^{\chi }{%
\delta v}+h_{5}\left( x^{k},v,\chi \right) \ ~^{\chi }{\delta y}\otimes
~^{\chi }{\delta y},  \notag \\
~^{\chi }\delta v &=&dv+w_{i}\left( x^{k},v,\chi \right) dx^{i},\ ~^{\chi
}\delta y=dy+n_{i}\left( x^{k},v,\chi \right) dx^{i},  \label{5ans5dr}
\end{eqnarray}%
for $g_{1}=\pm 1,$ with corresponding flows for N--adapted bases,
\begin{eqnarray}
~^{\chi }\mathbf{e}_{\alpha } &=&\mathbf{e}_{\alpha }(\chi )=\left( ~^{\chi }%
\mathbf{e}_{i}=\mathbf{e}_{i}(\chi )=\partial _{i}-N_{i}^{a}(u,\chi
)\partial _{a},e_{a}\right) ,  \label{5dder} \\
~^{\chi }\mathbf{e}^{\alpha } &=&\mathbf{e}^{\alpha }(\chi )=\left(
e^{i},~^{\chi }\mathbf{e}^{a}=\mathbf{e}^{a}(\chi )=dy^{a}+N_{i}^{a}(u,\chi
)dx^{i}\right)  \label{5ddif}
\end{eqnarray}%
defined by $N_{i}^{4}(u,\chi )=w_{i}\left( x^{k},v,\chi \right) $ and $%
N_{i}^{5}(u,\chi )=n_{i}\left( x^{k},v,\chi \right) .$ For any fixed value
of $\chi ,$ we may omit the Ricci flow parametric dependence.

The frames (\ref{5ddif}) satisfy certain nonholonomy (equivalently,
anholonomy) relations
\begin{equation}
\lbrack \mathbf{e}_{\alpha },\mathbf{e}_{\beta }]=\mathbf{e}_{\alpha }%
\mathbf{e}_{\beta }-\mathbf{e}_{\beta }\mathbf{e}_{\alpha }=W_{\alpha \beta
}^{\gamma }\mathbf{e}_{\gamma },  \label{5nanhrel}
\end{equation}%
with anholonomy coefficients
\begin{equation}
W_{ia}^{b}=\partial _{a}N_{i}^{b}\mbox{ and }W_{ji}^{a}=\Omega _{ij}^{a}=%
\mathbf{e}_{j}(N_{i}^{a})-\mathbf{e}_{j}(N_{i}^{a}).  \label{5anhncc}
\end{equation}%
A local basis is holonomic (for instance, the local coordinate basis) if $%
W_{\alpha \beta }^{\gamma }=0$ and integrable, i.e. it defines a fibred
structure, if the curvature of N--connection $\Omega _{ij}^{a}=0.$

We can elaborate on a N--anholonomic manifold $\mathbf{V}$ (i.e. on a
manifold provided with N--connection structure) a N--adapted tensor and
differential calculus if we decompose the geometric objects and basic
equations with respect to N--adapted bases (\ref{5ddif}) and (\ref{5dder})
and using the canonical d--connection $\widehat{\mathbf{D}}=\nabla +\mathbf{%
Z,}$ see, for instance, the formulas (A.17) in Ref. \cite{5nhrf04} for the
components of distorsion tensor $\mathbf{Z,}$ which contains certain
nontrivial torsion coefficients induced by ''off--diagonal'' $N_{i}^{a}.$
Two different linear connections, $\nabla $ and $\widehat{\mathbf{D}},$
define respectively two different Ricci tensors, $~_{\mid }R_{\alpha \beta }$
and $\widehat{\mathbf{R}}_{\alpha \beta }=[\widehat{R}_{ij},\widehat{R}_{ia},%
\widehat{R}_{bj},\widehat{S}_{ab}]$ (see (A.12) and related formulas in \cite%
{5nhrf01}). Even, in general, $\widehat{\mathbf{D}}\neq \nabla ,$ for
certain classes of ansatz and N--adapted frames, we can obtain, for some
nontrivial coefficients, relations of type $\widehat{\mathbf{R}}_{\alpha
\beta ~}=~_{\mid }R_{\alpha \beta }.$ This allows us to constrain some
classes of solutions of the Einstein and/ or Ricci flow equations
constructed for a more general linear connection $\widehat{\mathbf{D}}$ to
define also solutions for the Levi Civita connection $\nabla .$\footnote{%
we note that such equalities are obtained by deformation of the nonholonomic
structure on a manifold which change the transformation laws of tensors and
different linear connections}

For a 5D space initially provided with a diagonal ansatz (\ref{5dmetr}),
when $g_{ij}=diag[\pm 1,g_{2},g_{3}]$ and $g_{ab}=diag[g_{4},g_{5}],$ we
considered \cite{5vrf,5nhrf04} the nonholonomic (normalized) evolution
equations (parametrized by an ansatz (\ref{5ans5dr}))
\begin{eqnarray}
\frac{\partial }{\partial \chi }g_{ii} &=&-2\left[ \widehat{R}_{ii}-\lambda
g_{ii}\right] -h_{cc}\frac{\partial }{\partial \chi }(N_{i}^{c})^{2},
\label{5eq1} \\
\frac{\partial }{\partial \chi }h_{aa} &=&-2\ \left( \widehat{R}%
_{aa}-\lambda h_{aa}\right) ,\   \label{5eq2} \\
\ \widehat{R}_{\alpha \beta } &=&0\mbox{ for }\ \alpha \neq \beta ,
\label{5eq3}
\end{eqnarray}%
with the coefficients defined with respect to N--adapted frames (\ref{5dder}%
) and (\ref{5ddif}). This system of constrained (nonholonomic) evolution
equations in a particular case is related to families of metrics $\
^{\lambda }\mathbf{g}=\ ^{\lambda }\mathbf{g}_{\alpha \beta }\ \mathbf{e}%
^{\alpha }\otimes \mathbf{e}^{\beta }$ for nonholonomic Einstein spaces,
considered as solutions of
\begin{equation}
\widehat{\mathbf{R}}_{\alpha \beta }=\lambda \mathbf{g}_{\alpha \beta },
\label{5ep1b}
\end{equation}%
with effective cosmological constant $\lambda $ (in more general cases, we
can consider effective, locally anisotropically polarized cosmological
constants with dependencies on coordinates and $\chi :$ such solutions were
constructed in Refs. \cite{5vesnc,5vsgg}). For any solution of (\ref{5ep1b})
with nontrivial
\begin{equation*}
\mathbf{g}_{\alpha \beta }=[\pm
1,g_{2,3}(x^{2},x^{3}),h_{4,5}(x^{2},x^{3},v)]\mbox{\ and\ }%
w_{2,3}(x^{2},x^{3},v),n_{2,3}(x^{2},x^{3},v),
\end{equation*}%
we can consider nonholonomic Ricci flows of the horizontal metric
components, $g_{2,3}(x^{2},x^{3})\rightarrow g_{2,3}(x^{2},x^{3},\chi ),$
and of certain N--connection coefficients $n_{2,3}(x^{2},x^{3},v)\rightarrow
n_{2,3}(x^{2},x^{3},v,\chi ),$ constrained to satisfy the equation (\ref%
{5eq1}), i.e.%
\begin{equation}
\frac{\partial }{\partial \chi }\left[ g_{2,3}(x^{2},x^{3},\chi
)+h_{5}(x^{2},x^{3},v)~(n_{2,3}(x^{2},x^{3},v,\chi ))^{2}\right] =0.
\label{5eq1b}
\end{equation}%
Having constrained an integral variety of (\ref{5ep1b}) in order to have $%
\widehat{\mathbf{R}}_{\alpha \beta }=~_{\mid }R_{\alpha \beta }$ for certain
subclasses of solutions \footnote{%
this imposes certain additional restrictions on $n_{2,3}$ and $g_{2,3},$ see
discussions related to formulas (A.16)--(A.20) and explicit examples for
Ricci flows stated by constraints (49), or (74) in Ref. \cite{5nhrf04}); in
the next sections, we shall consider explicit examples}, the equations (\ref%
{5eq1b}) define evolutions of the geometric objects just for a family of
Levi Civita connections $~^{\chi }\nabla $.

Computing the components of the Ricci and Einstein tensors for the metric (%
\ref{5ans5dr}) (see details on tensors components' calculus in Refs. \cite%
{5vesnc,5vsgg}), one proves that the equations (\ref{5ep1b}) transform into
a parametric on $\chi $ system of partial differential equations:
\begin{eqnarray}
&&\widehat{R}_{2}^{2}=\widehat{R}_{3}^{3}(\chi )=\frac{1}{2g_{2}(\chi
)g_{3}(\chi )}[\frac{g_{2}^{\bullet }(\chi )g_{3}^{\bullet }(\chi )}{%
2g_{2}(\chi )}+  \label{5ep1a} \\
&& \qquad \frac{(g_{3}^{\bullet }(\chi ))^{2}}{2g_{3}(\chi )}-g_{3}^{\bullet
\bullet }(\chi )+\frac{g_{2}^{^{\prime }}(\chi )g_{3}^{^{\prime }}(\chi )}{%
2g_{3}(\chi )}+\frac{(g_{2}^{^{\prime }}(\chi ))^{2}}{2g_{2}(\chi )}%
-g_{2}^{^{\prime \prime }}(\chi )]=-\lambda ,  \notag \\
&&\widehat{S}_{4}^{4}=\widehat{S}_{5}^{5}(\chi )=\frac{1}{2h_{4}(\chi
)h_{5}(\chi )} \times  \notag \\
&&\qquad \left[ h_{5}^{\ast }(\chi )\left( \ln \sqrt{|h_{4}(\chi )h_{5}(\chi
)|}\right) ^{\ast }-h_{5}^{\ast \ast }(\chi )\right] =-\lambda ,
\label{5ep2a} \\
&&\widehat{R}_{4i}(\chi )=-w_{i}(\chi )\frac{\beta (\chi )}{2h_{5}(\chi )}-%
\frac{\alpha _{i}(\chi )}{2h_{5}(\chi )}=0,  \label{5ep3a} \\
&&\widehat{R}_{5i}(\chi )=-\frac{h_{5}(\chi )}{2h_{4}(\chi )}\left[
n_{i}^{\ast \ast }(\chi )+\gamma (\chi )n_{i}^{\ast }(\chi )\right] =0,
\label{5ep4a}
\end{eqnarray}%
where, for $h_{4,5}^{\ast }\neq 0,$%
\begin{eqnarray}
\alpha _{i} &=&h_{5}^{\ast }\partial _{i}\phi ,\ \beta =h_{5}^{\ast }\ \phi
^{\ast },\   \label{5coef} \\
\gamma &=&\frac{3h_{5}^{\ast }}{2h_{5}}-\frac{h_{4}^{\ast }}{h_{4}},~\phi
=\ln |\frac{h_{5}^{\ast }}{\sqrt{|h_{4}h_{5}|}}|,  \label{5coefa}
\end{eqnarray}%
when the necessary partial derivatives are written in the form $a^{\bullet
}=\partial a/\partial x^{2},$\ $a^{\prime }=\partial a/\partial x^{3},$\ $%
a^{\ast }=\partial a/\partial v.$ In the vacuum case, we shall put $\lambda
=0.$ Here we note that the dependence on $\chi $ can be considered both for
classes of functions and integrations constants and functions defining some
exact solutions of the Einstein equations or even for any general metrics on
a Rimann--Cartan manifold (provided with any compatible metric and linear
connection structures).

For an ansatz (\ref{5ans5dr}) with $g_{2}(x^{2},x^{3},\chi )=\epsilon
_{2}e^{\psi (x^{2},x^{3},\chi )}$ and $g_{2}(x^{2},x^{3},\chi )=\epsilon
_{3}e^{\psi (x^{2},x^{3},\chi )},$ we can restrict the solutions of the
system (\ref{5ep1a})--(\ref{5ep4a}) to define Ricci flows solutions with the
Levi Civita connection if the coefficients satisfy the conditions
\begin{eqnarray}
\epsilon _{2}\psi ^{\bullet \bullet }(\chi )+\epsilon _{3}\psi ^{^{\prime
\prime }}(\chi ) &=&\lambda  \notag \\
h_{5}^{\ast }\phi /h_{4}h_{5} &=&\lambda ,  \label{5ep2b} \\
w_{2}^{\prime }-w_{3}^{\bullet }+w_{3}w_{2}^{\ast }-w_{2}w_{3}^{\ast } &=&0,
\notag \\
n_{2}^{\prime }(\chi )-n_{3}^{\bullet }(\chi ) &=&0,  \notag
\end{eqnarray}%
for
\begin{equation}
w_{\widehat{i}}=\partial _{\widehat{i}}\phi /\phi ^{\ast },\mbox{\ where \ }%
\varphi =-\ln \left| \sqrt{|h_{4}h_{5}|}/|h_{5}^{\ast }|\right| ,
\label{5ep2c}
\end{equation}%
for $\widehat{i}=2,3,$ see formulas (49) in Ref. \cite{5nhrf04}.

\subsection{Five classes of primary metrics}

We introduce a list of 5D quadratic elements, defined by certain primary
metrics, which will be subjected to parametrized nonholonomic transforms in
order to generate new classes of exact solutions of the Einstein and Ricci
flow equations, i.e. of the system (\ref{5ep1a})--(\ref{5ep4a}) and (\ref%
{5ep1b}) with possible additional constraints in order to get geometric
evolutions in terms of the Levi Civita connection $\nabla .$

The first type quadratic element is taken
\begin{equation}
\delta s_{[1]}^{2}=\epsilon _{1}d\varkappa ^{2}-d\xi ^{2}-r^{2}(\xi )\
d\vartheta ^{2}-r^{2}(\xi )\sin ^{2}\vartheta \ d\varphi ^{2}+\varpi
^{2}(\xi )\ dt^{2},  \label{5aux1}
\end{equation}%
where the local coordinates and nontrivial metric coefficients are
parametriz\-ed in the form%
\begin{eqnarray}
x^{1} &=&\varkappa ,x^{2}=\xi ,x^{3}=\vartheta ,y^{4}=\varphi ,y^{5}=t,
\label{5aux1p} \\
\check{g}_{1} &=&\epsilon _{1}=\pm 1,\ \check{g}_{2}=-1,\ \check{g}%
_{3}=-r^{2}(\xi ),\ \check{h}_{4}=-r^{2}(\xi )\sin ^{2}\vartheta ,\ \check{h}%
_{5}=\varpi ^{2}(\xi ),  \notag
\end{eqnarray}%
for
\begin{equation*}
\xi =\int dr\ \left| 1-\frac{2\mu }{r}+\frac{\varepsilon }{r^{2}}\right|
^{1/2}\mbox{\ and\ }\varpi ^{2}(r)=1-\frac{2\mu }{r}+\frac{\varepsilon }{%
r^{2}}.
\end{equation*}%
For the constants $\varepsilon \rightarrow 0$ and $\mu $ being a point mass,
the element (\ref{5aux1}) defines just a trivial embedding into 5D (with
extra dimension coordinate $\varkappa )$ of the Schwarzschild solution
written in spacetime spherical coordinates $(r,\vartheta ,\varphi ,t).$%
\footnote{%
For simplicity, we consider only the case of vacuum solutions, not analyzing
a more general possibility when $\varepsilon =e^{2}$ is related to the
electric charge for the Reissner--Nordstr\"{o}m metric (see, for example, %
\cite{5heu}). In our further considerations, we shall treat $\varepsilon $
as a small parameter, for instance, defining a small deformation of a circle
into an ellipse (eccentricity).}

The second quadratic element is%
\begin{equation}
\delta s_{[2]}^{2}=-r_{g}^{2}\ d\varphi ^{2}-r_{g}^{2}\ d\check{\vartheta}%
^{2}+\check{g}_{3}(\check{\vartheta})\ d\check{\xi}^{2}+\epsilon _{1}\ d\chi
^{2}+\check{h}_{5}\ (\xi ,\check{\vartheta})\ dt^{2},  \label{5aux2}
\end{equation}%
where the local coordinates are
\begin{equation*}
x^{1}=\varphi ,x^{2}=\check{\vartheta},x^{3}=\check{\xi},y^{4}=\chi ,y^{5}=t,
\end{equation*}%
for
\begin{equation*}
d\check{\vartheta}=d\vartheta /\sin \vartheta ,\ d\check{\xi}=dr/r\sqrt{%
|1-2\mu /r+\varepsilon /r^{2}|},
\end{equation*}%
and the Schwarzschild radius of a point mass $\mu $ is defined $%
r_{g}=2G_{[4]}\mu /c^{2},$ where $G_{[4]}$ is the 4D Newton constant and $c$
is the light velocity. The nontrivial metric coefficients in (\ref{5aux2})
are parametrized%
\begin{eqnarray}
\check{g}_{1} &=&-r_{g}^{2},\ \check{g}_{2}=-r_{g}^{2},\ \check{g}%
_{3}=-1/\sin ^{2}\vartheta ,  \label{5aux2p} \\
\ \check{h}_{4} &=&\epsilon _{1},\ \check{h}_{5}=\left[ 1-2\mu
/r+\varepsilon /r^{2}\right] /r^{2}\sin ^{2}\vartheta .  \notag
\end{eqnarray}%
The quadratic element defined by (\ref{5aux2}) and (\ref{5aux2p}) is a
trivial embedding into 5D of the Schwarzschild quadratic element multiplied
to the conformal factor $\left( r\sin \vartheta /r_{g}\right) ^{2}.$ We
emphasize that this metric is not a solution of the Einstein or Ricci flow
equations but it will be used in order to construct parametrized
nonholonomic deformations to such solutions.

We shall use a quadratic element when the time coordinate is considered to
be ''anisotropic'',
\begin{equation}
\delta s_{[3]}^{2}=-r_{g}^{2}\ d\varphi ^{2}-r_{g}^{2}\ d\check{\vartheta}%
^{2}+\check{g}_{3}(\check{\vartheta})\ d\check{\xi}^{2}+\check{h}_{4}\ (\xi ,%
\check{\vartheta})\ dt^{2}+\epsilon _{1}\ d\varkappa ^{2}  \label{5aux3}
\end{equation}%
where the local coordinates are
\begin{equation*}
x^{1}=\varphi ,\ x^{2}=\check{\vartheta},\ x^{3}=\check{\xi},\ y^{4}=t,\
y^{5}=\varkappa ,
\end{equation*}%
and the nontrivial metric coefficients are parametrized%
\begin{eqnarray}
\check{g}_{1} &=&-r_{g}^{2},\ \check{g}_{2}=-r_{g}^{2},\ \check{g}%
_{3}=-1/\sin ^{2}\vartheta ,\   \label{5aux3p} \\
\check{h}_{4} &=&\left[ 1-2\mu /r+\varepsilon /r^{2}\right] /r^{2}\sin
^{2}\vartheta ,\ \check{h}_{5}=\epsilon _{1}.  \notag
\end{eqnarray}%
The formulas (\ref{5aux3}) and (\ref{5aux3p}) are respective
reparametrizations of (\ref{5aux2}) and (\ref{5aux2p}) when the $4$th and $5$%
th coordinates are inverted. Such metrics will be used for constructing new
classes of exact solutions in 5D with explicit dependence on time like
coordinate.

The forth quadratic element is introduced by inverting the 4th and 5th
coordinates in (\ref{5aux1})
\begin{equation}
\delta s_{[4]}^{2}=\epsilon _{1}d\varkappa ^{2}-d\xi ^{2}-r^{2}(\xi )\
d\vartheta ^{2}+\varpi ^{2}(\xi )\ dt^{2}-r^{2}(\xi )\sin ^{2}\vartheta \
d\varphi ^{2}  \label{5aux4}
\end{equation}%
where the local coordinates and nontrivial metric coefficients are
parametriz\-ed in the form%
\begin{eqnarray}
x^{1} &=&\varkappa ,x^{2}=\xi ,x^{3}=\vartheta ,y^{4}=t,y^{5}=\varphi ,
\label{5aux4p} \\
\check{g}_{1} &=&\epsilon _{1}=\pm 1,\ \check{g}_{2}=-1,\ \check{g}%
_{3}=-r^{2}(\xi ),\ \check{h}_{4}=\varpi ^{2}(\xi ),\ \check{h}%
_{5}=-r^{2}(\xi )\sin ^{2}\vartheta .  \notag
\end{eqnarray}%
Such metrics can be used for constructing exact solutions in 4D gravity and
Ricci flows with anisotropic dependence on time coordinate.

Finally, we consider
\begin{equation}
\delta s_{[5]}^{2}=\epsilon _{1}\ d\varkappa ^{2}-dx^{2}-dy^{2}-2\kappa
(x,y,p)\ dp^{2}+\ dv^{2}/8\kappa (x,y,p),  \label{5aux5}
\end{equation}%
where the local coordinates are
\begin{equation*}
\ x^{1}=\varkappa ,\ x^{2}=x,\ x^{3}=y,\ y^{4}=p,\ y^{5}=v,
\end{equation*}%
and the nontrivial metric coefficients are parametrized%
\begin{eqnarray}
\check{g}_{1} &=&\epsilon _{1}=\pm 1,\ \check{g}_{2}=-1,\ \check{g}_{3}=-1,\
\label{5aux5p} \\
\check{h}_{4} &=&-2\kappa (x,y,p),\ \check{h}_{5}=1/\ 8\ \kappa (x,y,p).
\notag
\end{eqnarray}%
The metric (\ref{5aux5}) is a trivial embedding into 5D of the vacuum
solution of the Einstein equation defining pp--waves \cite{5peres} for any $%
\kappa (x,y,p)$ solving
\begin{equation*}
\kappa _{xx}+\kappa _{yy}=0,
\end{equation*}%
with $p=z+t$ and $v=z-t,$ where $(x,y,z)$ are usual Cartesian coordinates
and $t$ is the time like coordinates. The simplest explicit examples of such
solutions are
\begin{equation*}
\kappa =(x^{2}-y^{2})\sin p,
\end{equation*}%
defining a plane monochromatic wave, or
\begin{eqnarray*}
\kappa &=&\frac{xy}{\left( x^{2}+y^{2}\right) ^{2}\exp \left[ p_{0}^{2}-p^{2}%
\right] },\mbox{ for }|p|<p_{0}; \\
&=&0,\mbox{ for }|p|\geq p_{0},
\end{eqnarray*}%
defining a wave packet travelling with unit velocity in the negative $z$
direction.

\section{Solitonic pp--Waves and String Torsion}

Pp--wave solutions are intensively exploited for elaborating string models
with nontrivial backgrounds \cite{5strpp1,5strpp2,5strpp3}. A special
interest for pp--waves in general relativity is related to the fact that any
solution in this theory can be approximated by a pp--wave in vicinity of
horizons. Such solutions can be generalized by introducing nonlinear
interactions with solitonic waves \cite{5vs1,5gravsolit,5bv,5vhep,5vp} and
nonzero sources with nonhomogeneous cosmological constant induced by an
ansatz for the antisymmetric tensor fields of third rank, see Appendix. A
very important property of such nonlinear wave solutions is that they
possess nontrivial limits defining new classes of generic off--diagonal
vacuum Einstein spacetimes and can be generalized for Ricci flows induced by
evolutions of N--connections.

In this section, we use an ansatz of type (\ref{5ans5dr}),
\begin{eqnarray}
\delta s_{[5]}^{2} &=&\epsilon _{1}\ d\varkappa ^{2}-e^{\psi (x,y,\chi
)}\left( dx^{2}+dy^{2}\right)  \label{5sol2} \\
&&-2\kappa (x,y,p)\ \eta _{4}(x,y,p)\delta p^{2}+\ \frac{\eta _{5}(x,y,p)}{%
8\kappa (x,y,p)}\delta v^{2}  \notag \\
\delta p &=&dp+w_{2}(x,y,p)dx+w_{3}(x,y,p)dy,\   \notag \\
\delta v &=&dv+n_{2}(x,y,p,\chi )dx+n_{3}(x,y,p,\chi )dy  \notag
\end{eqnarray}%
where the local coordinates are
\begin{equation*}
\ x^{1}=\varkappa ,\ x^{2}=x,\ x^{3}=y,\ y^{4}=p,\ y^{5}=v,
\end{equation*}%
and the nontrivial metric coefficients and polarizations are parametrized%
\begin{eqnarray*}
\check{g}_{1} &=&\epsilon _{1}=\pm 1,\ \check{g}_{2}=-1,\ \check{g}_{3}=-1,
\\
\check{h}_{4}&=&-2\kappa (x,y,p),\ \check{h}_{5}=1/\ 8\kappa (x,y,p), \\
\eta _{1} &=&1,g_{\alpha }=\eta _{\alpha }\check{g}_{\alpha }.
\end{eqnarray*}%
For trivial polarizations $\eta _{\alpha }=1$ and $w_{2,3}=0,$ $n_{2,3}=0,$
the metric (\ref{5sol2}) is just the pp--wave solution (\ref{5aux5}).

\subsection{Ricci flows of solitonic pp--wave solutions in string gravity}

Our aim is to define such nontrivial values of polarization functions when $%
\eta _{5}(x,y,p)$ is defined by a 3D soliton $\eta (x,y,p),$ for instance,
as a solution of solitonic equation
\begin{equation}
\eta ^{\bullet \bullet }+\epsilon (\eta ^{\prime }+6\eta \ \eta ^{\ast
}+\eta ^{\ast \ast \ast })^{\ast }=0,\ \epsilon =\pm 1,  \label{5solit1}
\end{equation}%
and $\eta _{2}=\eta _{3}=e^{\psi (x,y,\chi )}$ is a family solutions of (\ref%
{5ep1a}) transformed into
\begin{equation}
\psi ^{\bullet \bullet }(\chi )+\psi ^{\prime \prime }(\chi )=\frac{\lambda
_{H}^{2}}{2}.  \label{5lapl}
\end{equation}%
The solitonic deformations of the pp--wave metric will define exact
solutions in string gravity with $H$--fields, see in Appendix the equations (%
\ref{5c01}) and (\ref{5c02}) for the string torsion ansatz (\ref{5ansh}),
when with $\lambda =\lambda _{H}.$\footnote{%
as a matter of principle we can consider\ that $\phi $ is a solution of any
3D solitonic, or other, nonlinear wave equation.}

Introducing the\ above stated data for the ansatz (\ref{5sol2}) into the
equation (\ref{5ep2a}),\footnote{%
such solutions can be constructed in general form (see, in details, the
formulas (26)--(28) in Ref. \cite{5nhrf04}, for corresponding
reparametrizations)} we get two equations relating $h_{4}=\eta _{4}\check{h}%
_{4}$ and $h_{5}=\eta _{5}\check{h}_{5},$
\begin{equation}
\eta _{5}=8\ \kappa (x,y,p)\left[ h_{5[0]}(x,y)+\frac{1}{2\lambda _{H}^{2}}%
e^{2\eta (x,y,p)}\right]  \label{5sol2h5}
\end{equation}%
and
\begin{equation}
|\eta _{4}|=\frac{e^{-2\phi (x,y,p)}}{2\kappa ^{2}(x,y,p)}\left[ \left(
\sqrt{|\eta _{5}|}\right) ^{\ast }\right] ^{2},  \label{5sol2h4}
\end{equation}%
where $h_{5[0]}(x,y)$ is an integration function. Having defined the
coefficients $h_{a},$ we can solve the equations (\ref{5ep3a}) and (\ref%
{5ep4a}) expressing the coefficients (\ref{5coef}) and (\ref{5coefa})
through $\eta _{4}$ and $\eta _{5}$ defined by pp-- and solitonic waves as
in (\ref{5sol2h4}) and (\ref{5sol2h5}). The corresponding solutions are
\begin{equation}
w_{1}=0,w_{2}=\left( \phi ^{\ast }\right) ^{-1}\partial _{x}\phi
,w_{3}=\left( \phi ^{\ast }\right) ^{-1}\partial _{x}\phi ,  \label{5sol2w}
\end{equation}%
for $\phi ^{\ast }=\partial \phi /\partial p,$ see formulas (\ref{5ep2c})
and
\begin{equation}
n_{1}=0,n_{2,3}=n_{2,3}^{[0]}(x,y,\chi )+n_{2,3}^{[1]}(x,y,\chi )\int \left|
\eta _{4}\eta _{5}^{-3/2}\right| dp,  \label{5sol2na}
\end{equation}%
where $n_{2,3}^{[0]}(x,y,\chi )$ and $n_{2,3}^{[1]}(x,y,\chi )$ are
integration functions, restricted to satisfy the conditions (\ref{5eq1b}),%
\begin{eqnarray}
&&\frac{\partial }{\partial \chi }[-e^{\psi (x,y,\chi )}+\eta _{5}(x,y,p)~%
\check{h}_{5}(x,y,p)~(n_{2,3}^{[0]}(x,y,\chi )  \label{5eq1c} \\
&&+n_{2,3}^{[1]}(x,y,\chi )\int \left| \eta _{4}(x,y,p)\eta
_{5}^{-3/2}(x,y,p)\right| dp)^{2}]=0.  \notag
\end{eqnarray}

We note that the ansatz (\ref{5sol2}), without dependence on $\chi $ and
with the coefficients computed following the equations and formulas (\ref%
{5lapl}), (\ref{5sol2h4}), (\ref{5sol2h5}), (\ref{5sol2w}) and (\ref{5sol2na}%
), defines a class of exact solutions (depending on integration functions)
of gravitational field equations in string gravity with $H$--field. For
corresponding families of coefficients evolving on $\chi $ and constrained
to satisfy the conditions (\ref{5eq1c}) we get solutions of nonholonomic
Ricci flow equations (\ref{5eq1})--(\ref{5eq3}) normalized by the effective
constant $\lambda _{H}$ induced from string gravity.

Putting the above stated functions $\psi ,k,\phi $ and $\eta _{5}$ and
respective integration functions into the corresponding ansatz, we define a
class of evolution and/or gravity field solutions,
\begin{eqnarray}
\delta s_{[sol2]}^{2} &=&\epsilon _{1}\ d\varkappa ^{2}-e^{\psi (\chi
)}\left( dx^{2}+dy^{2}\right)  \notag \\
&&+\ \frac{\eta _{5}}{8\kappa }\delta p^{2}-\kappa ^{-1}\ e^{-2\phi }\left[
\left( \sqrt{|\eta _{5}|}\right) ^{\ast }\right] ^{2}\delta \nu ^{2}(\chi ),
\notag \\
\delta p &=&dp+\left( \phi ^{\ast }\right) ^{-1}\partial _{x}\phi \
dx+\left( \phi ^{\ast }\right) ^{-1}\partial _{y}\phi \ dy,\   \label{5sol2a}
\\
\delta v(\chi ) &=&dv+\left\{ n_{2}^{[0]}(\chi )+\widehat{n}_{2}^{[1]}(\chi
)\int k^{-1}e^{2\phi }\left[ \left( \left| \eta _{5}\right| ^{-1/4}\right)
^{\ast }\right] ^{2}dp\right\} dx  \notag \\
&&+\left\{ n_{3}^{[0]}(\chi )+\widehat{n}_{3}^{[1]}(\chi )\int
k^{-1}e^{2\phi }\left[ \left( \left| \eta _{5}\right| ^{-1/4}\right) ^{\ast }%
\right] ^{2}dp\right\} dy,  \notag
\end{eqnarray}%
where some constants and multiples depending on $x$ and $y$ are included
into $\widehat{n}_{2,3}^{[1]}(x,y,\chi )$ and we emphasize the dependence of
coefficients on Ricci flow parameter $\chi .$ Such families of generic
off--diagonal metrics posses induced both nonholonomically and from string
gravity torsion coefficients for the canonical d--connection (we omit
explicit formulas for the nontrivial components which can be computed by
introducing the coefficients of our ansatz into (\ref{5torsa})). This class
of solutions describes nonlinear interactions of pp--waves and 3D solutions
in string gravity in Ricci flow theory.

The term $\epsilon _{1}\ d\varkappa ^{2}$ can be eliminated in order to
describe only 4D configurations. Nevertheless, in this case, there is not a
smooth limit of such 4D solutions for $\lambda _{H}^{2}\rightarrow 0$ to
those in general relativity, see the second singular term in (\ref{5sol2h5}%
), proportional to $1/\lambda _{H}^{2}.$

Finally, note that explicit values for the integration functions and
constants can be defined (for a fixed system of reference and coordinates)
from certain initial value and boundary conditions. In this work, we shall
analyze the properties of the derived classes of solutions and their
multi--parametric transforms and geometric flows working with general forms
of generation and integration functions.

\subsection{Solitonic pp--waves in vacuum Einstein gravity and Ricci flows}

\label{5ssaux2} In this section, we show how the anholonomic frame method
can be used for constructing 4D metrics induced by nonlinear pp--waves and
solitonic interactions for vanishing sources and the Levi Civita connection.
For an ansatz of type (\ref{5sol2}), we write
\begin{equation*}
\eta _{5}=5\kappa b^{2}\mbox{ and }\eta _{4}=h_{0}^{2}(b^{\ast
})^{2}/2\kappa .
\end{equation*}%
A 3D solitonic solution can be generated if $b$ is subjected to the
condition to solve a solitonic equation, for instance, of type (\ref{5solit1}%
), or other nonlinear wave configuration. We chose a parametrization when
\begin{equation*}
b(x,y,p)=\breve{b}(x,y)q(p)k(p),
\end{equation*}%
for any $\breve{b}(x,y)$ and any pp--wave $\kappa (x,y,p)=\breve{\kappa}%
(x,y)k(p)$ (we can take $\breve{b}=\breve{\kappa}),$ where $q(p)=4\tan
^{-1}(e^{\pm p})$ is the solution of ''one dimensional'' solitonic equation
\begin{equation}
q^{\ast \ast }=\sin q.  \label{5sol1d}
\end{equation}%
In this case,
\begin{equation}
w_{2}=\left[ (\ln |qk|)^{\ast }\right] ^{-1}\partial _{x}\ln |\breve{b}|%
\mbox{ and }w_{3}=\left[ (\ln |qk|)^{\ast }\right] ^{-1}\partial _{y}\ln |%
\breve{b}|.  \label{5aux5aa}
\end{equation}%
The final step in constructing such vacuum Einstein solutions is to chose
any two functions $n_{2,3}(x,y)$ satisfying the conditions $n_{2}^{\ast
}=n_{3}^{\ast }=0$ \ and $n_{2}^{\prime }-n_{3}^{\bullet }=0$ which are
necessary for Riemann foliated structures with the Levi Civita connection,
see discussion of formulas (42) and (43) in Ref. \cite{5nhrf04} and
conditions (\ref{5ep2b}). This mean that in the integrals of type (\ref%
{5sol2na}) we shall fix the integration functions $n_{2,3}^{[1]}=0$ but take
such $n_{2,3}^{[0]}(x,y)$ satisfying $(n_{2}^{[0]})^{\prime
}-(n_{3}^{[0]})^{\bullet }=0.$

We can consider a trivial solution of (\ref{5ep1a}), i.e. of (\ref{5lapl})
with $\lambda =\lambda _{H}=0.$

Summarizing the results, we obtain the 4D vacuum metric
\begin{eqnarray}
\delta s_{[sol2a]}^{2} &=&-\left( dx^{2}+dy^{2}\right) -h_{0}^{2}\breve{b}%
^{2}[(qk)^{\ast }]^{2}\delta p^{2}+\breve{b}^{2}(qk)^{2}\delta v^{2},  \notag
\\
\delta p &=&dp+\left[ (\ln |qk|)^{\ast }\right] ^{-1}\partial _{x}\ln |%
\breve{b}|\ dx+\left[ (\ln |qk|)^{\ast }\right] ^{-1}\partial _{y}\ln |%
\breve{b}|\ dy,\   \notag \\
\delta v &=&dv+n_{2}^{[0]}dx+n_{3}^{[0]}dy,  \label{5sol2b}
\end{eqnarray}%
defining nonlinear gravitational interactions of a pp--wave $\kappa =\breve{%
\kappa}k$ and a soliton $q,$ depending on certain type of integration
functions and constants stated above. Such vacuum Einstein metrics can be
generated in a similar form for 3D or 2D solitons but the constructions will
be more cumbersome and for non--explicit functions, see a number of similar
solutions in Refs. \cite{5vs1,5vsgg}.

Now, we generalize the ansatz (\ref{5sol2b}) in a form describing normalized
Ricci flows of the mentioned type vacuum solutions extended for a prescribed
constant $\lambda $ necessary for normalization. We chose
\begin{eqnarray}
\delta s_{[sol2a]}^{2} &=&-\left( dx^{2}+dy^{2}\right) -h_{0}^{2}\breve{b}%
^{2}(\chi )[(qk)^{\ast }]^{2}\delta p^{2}+\breve{b}^{2}(\chi )(qk)^{2}\delta
v^{2},  \notag \\
\delta p &=&dp+\left[ (\ln |qk|)^{\ast }\right] ^{-1}\partial _{x}\ln |%
\breve{b}|\ dx+\left[ (\ln |qk|)^{\ast }\right] ^{-1}\partial _{y}\ln |%
\breve{b}|\ dy,\   \notag \\
\delta v &=&dv+n_{2}^{[0]}(\chi )dx+n_{3}^{[0]}(\chi )dy,  \label{5sol2bf}
\end{eqnarray}%
where we introduced the parametric dependence on $\chi ,$
\begin{equation*}
b(x,y,p,\chi )=\breve{b}(x,y,\chi )q(p)k(p)
\end{equation*}%
which allows us to use the same formulas (\ref{5aux5aa}) for $w_{3,4}$ not
depending on $\chi .$ The values $\breve{b}^{2}(\chi )$ and $%
n_{2}^{[0]}(\chi )$ are constrained to be solutions of
\begin{equation}
\frac{\partial }{\partial \chi }\left[ \breve{b}^{2}(n_{2,3}^{[0]})^{2}%
\right] =-2\lambda \mbox{ and }\frac{\partial }{\partial \chi }\breve{b}%
^{2}=2\lambda \breve{b}^{2}\   \label{5const5a}
\end{equation}%
in order to solve, respectively, the equations (\ref{5eq1}) and (\ref{5eq2}%
). As a matter of principle, we can consider a flow dependence as a factor $%
\psi (\lambda )$ before $\left( dx^{2}+dy^{2}\right) ,$ i.e. flows of the
h--components of metrics which will generalize the ansatz (\ref{5sol2bf})
and constraints (\ref{5const5a}). For simplicity, we have chosen a minimal
extension of vacuum Einstein solutions in order to describe nonholonomic
flows of the v--components of metrics adapted to the flows of N--connection
coefficients $n_{2,3}^{[0]}(\chi ).$ Such nonholonomic constraints on metric
coefficients define Ricci flows of families of vacuum Einstein solutions
defined by nonlinear interactions of a 3D soliton and a pp--wave.

\section{Parametric Transforms and Flows and Solitonic pp--Wa\-ve Solutions}

There are different possibilities to apply parametric and frame transforms
and define Ricci flows and nonholonomic deformations of geometric objects.
The first one is to perform a parametric transform of a vacuum solution and
then to deform it nonholonomically in order to generate pp--wave solitonic
interactions. In the second case, we can subject an already nonholonomically
generated solution of type (\ref{5sol2b}) to a one parameter transforms.
Finally, in the third case, we can derive two parameter families of
nonholonomic soliton pp--wave interactions. For simplicity, Ricci flows will
be considered after certain classes of exact solutions of field equations
will have been constructed.

\subsection{Flows of solitonic pp--waves generated \newline
by parametric tran\-sforms}

Let us consider the metric
\begin{equation}
\delta s_{[5a]}^{2}=-dx^{2}-dy^{2}-2\breve{\kappa}(x,y)\ dp^{2}+\ dv^{2}/8%
\breve{\kappa}(x,y)  \label{5auxpw}
\end{equation}%
which is a particular 4D case\ of (\ref{5aux5}) when $\kappa
(x,y,p)\rightarrow \breve{\kappa}(x,y).$ It is easy to show that the
nontrivial Ricci components $R_{\alpha \beta }$ for the Levi Civita
connection are proportional to $\breve{\kappa}^{\bullet \bullet }+\breve{%
\kappa}^{\prime \prime }$ and the non--vanishing components of the curvature
tensor $R_{\alpha \beta \gamma \delta }$ are of type $R_{a1b1}\simeq
R_{a2b2}\simeq \sqrt{\left( \breve{\kappa}^{\bullet \bullet }\right)
^{2}+\left( \breve{\kappa}^{\bullet \prime }\right) ^{2}}.$ So, any function
$\breve{\kappa}$ solving the equation $\breve{\kappa}^{\bullet \bullet }+%
\breve{\kappa}^{\prime \prime }=0$ but with $\left( \breve{\kappa}^{\bullet
\bullet }\right) ^{2}+\left( \breve{\kappa}^{\bullet \prime }\right)
^{2}\neq 0$ defines a vacuum solution of the Einstein equations. In the
simplest case, we can take $\breve{\kappa}=x^{2}-y^{2}$ or $\breve{\kappa}%
=xy/\sqrt{x^{2}+y^{2}}$ like it was suggested in the original work \cite%
{5peres}, but for the metric (\ref{5auxpw}) we do not consider any multiple $%
q(p)$ depending on $p.$

Subjecting the metric (\ref{5auxpw}) to a parametric transform, we get an
off--diagonal metric of type
\begin{eqnarray}
\delta s_{[2p]}^{2} &=&-\eta _{2}(x,y,\theta )dx^{2}+\eta _{3}(x,y,\theta
)dy^{2}  \label{5auxpwa} \\
&&-2\breve{\kappa}(x,y)\ \eta _{4}(x,y,\theta ,\chi )\delta p^{2}+\ \frac{%
\eta _{5}(x,y,\theta ,\chi )}{8\breve{\kappa}(x,y)}\delta v^{2},  \notag \\
\delta p &=&dp+w_{2}(x,y,\theta )dx+w_{3}(x,y,\theta )dy,\   \notag \\
\delta v &=&dv+n_{2}(x,y,\theta ,\chi )dx+n_{3}(x,y,\theta ,\chi )dy  \notag
\end{eqnarray}%
which may define Ricci flows, or vacuum solutions of the Einstein equations,
if the coefficients are restricted to satisfy the necessary conditions. Such
parametric transforms consist a particular case of frame transforms when the
coefficients $g_{\underline{\alpha }\underline{\beta }}$ are defined by the
coefficients of (\ref{5auxpw}) and $_{\shortmid }^{\circ }\widetilde{g}%
_{\alpha \beta }$ are given by the coefficients (\ref{5auxpwa}). The
polarizations $\eta _{\widehat{\alpha }}(x,y,\theta ,\chi )$ and
N--connection coefficients $w_{\widehat{i}}(x,y,\theta )$ and $n_{\widehat{i}%
}(x,y,\theta ,\chi )$ determine the coefficients of the matrix of
parametric, or Geroch, transforms (for details on nonholonomic
generalizations and Geroch equations, see section 4.2 and Appendix B in Ref. %
\cite{5nhrf04} and sections 2.2 and 3 in Ref. \cite{5vnhes}; we note that in
this work we have an additional to $\theta $ Ricci flow parameter $\chi ).$

Considering that $\eta _{2}\neq 0,$\footnote{$\eta _{2}\rightarrow 1$ and $%
\eta _{3}\rightarrow 1$ for infinitesimal parametric transforms} we multiply
(\ref{5auxpwa}) on conformal factor $\left( \eta _{2}\right) ^{-1}$ and
redefining the coefficients as $\breve{\eta}_{3}=\eta _{3}/\eta _{2},\breve{%
\eta}_{a}=\eta _{a}/\eta _{2},$ $\breve{w}_{a}=w_{a}$ and $\breve{n}%
_{a}=n_{a},$ for $\hat{\imath}=2,3$ and $a=4,5,$ we obtain
\begin{eqnarray}
\delta s_{[2a]}^{2} &=&-dx^{2}+\breve{\eta}_{3}(x,y,\theta )dy^{2}
\label{5auxpwb} \\
&&-2\breve{\kappa}(x,y)\ \breve{\eta}_{4}(x,y,\theta ,\chi )\delta p^{2}+\
\frac{\breve{\eta}_{5}(x,y,\theta ,\chi )}{8\breve{\kappa}(x,y)}\delta v^{2}
\notag \\
\delta p &=&dp+\breve{w}_{2}(x,y,\theta )dx+\breve{w}_{3}(x,y,\theta )dy,\
\notag \\
\delta v &=&dv+\breve{n}_{2}(x,y,\theta ,\chi )dx+\breve{n}_{3}(x,y,\theta
,\chi )dy  \notag
\end{eqnarray}%
which is not an exact solution but can be nonholonomically deformed into
exact vacuum solutions by multiplying on additional polarization parameters.
Firstly, we first introduce the polarization $\eta _{2}=\exp \psi
(x,y,\theta ,\chi )$ when $\eta _{3}=\breve{\eta}_{3}=-\exp \psi (x,y,\theta
,\chi )$ are defined as families of solutions of $\psi ^{\bullet \bullet
}(\chi )+\psi ^{\prime \prime }(\chi )=\lambda .$ Then, secondly, we
redefine $\breve{\eta}_{a}\rightarrow \eta _{a}(x,y,p,\chi )$ (for instance,
multiplying on additional multiples) by introducing additional dependencies
on ''anisotropic'' coordinate $p$ such a way when the ansatz (\ref{5auxpwb})
transform into
\begin{eqnarray}
\delta s_{[2a]}^{2} &=&-e^{\psi (x,y,\theta ,\chi )}\left(
dx^{2}+dy^{2}\right)  \label{5auxpwc} \\
&&-2\breve{\kappa}(x,y)k(p)\ \eta _{4}(x,y,p,\theta )\delta p^{2}+\ \frac{%
\eta _{5}(x,y,p,\theta )}{8\breve{\kappa}(x,y)k(p)}\delta v^{2}  \notag \\
\delta p &=&dp+w_{2}(x,y,p,\theta )dx+w_{3}(x,y,p,\theta )dy,\   \notag \\
\delta v &=&dv+n_{2}(x,y,\theta ,\chi )dx+n_{3}(x,y,\theta ,\chi )dy.  \notag
\end{eqnarray}%
The ''simplest'' Ricci flow solutions induced by flows of the h--metric and
N--connection coefficients are
\begin{equation}
w_{1}=0,w_{2}(\theta )=\left( \phi ^{\ast }\right) ^{-1}\partial _{x}\phi
,w_{3}(\theta )=\left( \phi ^{\ast }\right) ^{-1}\partial _{x}\phi ,
\label{5sol2wa}
\end{equation}%
for $\phi ^{\ast }=\partial \phi /\partial p,$ see formulas (\ref{5ep2c}),
and
\begin{equation}
n_{2,3}=n_{2,3}^{[0]}(x,y,\theta ,\chi )+n_{2,3}^{[1]}(x,y,\theta ,\chi
)\int \left| \eta _{4}(\theta )\eta _{5}^{-3/2}(\theta )\right| dp,
\label{5sol2naa}
\end{equation}%
where $n_{2,3}^{[0]}(x,y,,\theta ,\chi )$ and $n_{2,3}^{[1]}(x,y,,\theta
,\chi )$ are constrained as (\ref{5eq1b}),%
\begin{eqnarray}
&&\frac{\partial }{\partial \chi }[-e^{\psi (x,y,\theta ,\chi )}+\eta
_{5}(x,y,p,\theta )~\check{h}_{5}(x,y,p)~(n_{2,3}^{[0]}(x,y,\theta ,\chi )
\label{5eq1ca} \\
&&+n_{2,3}^{[1]}(x,y,\theta ,\chi )\int \left| \eta _{4}(x,y,p,\theta )\eta
_{5}^{-3/2}(x,y,p,\theta )\right| dp)^{2}]=0.  \notag
\end{eqnarray}%
The difference of formulas (\ref{5sol2wa}), (\ref{5sol2naa}) and (\ref%
{5eq1ca}) and respective formulas (\ref{5sol2w}), (\ref{5sol2na}) and (\ref%
{5eq1c}) is that the set\ of coefficients defining the nonhlonomic Ricci
flow of metrics (\ref{5auxpwc}) depend on a free parameter $\theta $
associated to some 'primary' Killing symmetries like it was considered by
Geroch \cite{5geroch1}. The analogy with Geroch's (parametric) transforms is
more complete if we do not consider dependencies on $\chi $ and take the
limit $\lambda \rightarrow 0$ which generates families, on $\theta ,$ of
vacuum Einstein solutions, see formula (105) in Ref. \cite{5vnhes}.

In order to define Ricci flows for the Levi Civita connection, with $g_{4}=-2%
\breve{\kappa}k\eta _{4}$ and $g_{5}=\eta _{5}/8\breve{\kappa}k$ from (\ref%
{5auxpwc}), the coefficients of this metric must solve the conditions (\ref%
{5ep2b}), when the coordinates are parametrized $x^{2}=x,x^{3}=y,y^{4}=p$
and $y^{5}=v.$ This describes both parametric nonholonomic transform and
Ricci flows of a metric (\ref{5auxpw}) to a family of evolution/ field exact
solutions depending on parameter $\theta $ and defining nonlinear
superpositions of pp--waves $\kappa =\breve{\kappa}(x,y)k(p).$

It is possible to introduce solitonic waves into the metric (\ref{5auxpwc}).
For instance, we can take $\eta _{5}(x,y,p,\theta )\sim q(p),$ where $q(p)$
is a solution of solitonic equation (\ref{5sol1d}). We obtain nonholonomic
Ricci flows of a family of Einstein metrics labelled by parameter $\theta $
and defining nonlinear interactions of pp--waves and one--dimensional
solitons. Such solutions with prescribed $\psi =0$ can be parametrized in a
form very similar to the ansatz (\ref{5sol2b}).

\subsection{Parametrized transforms and flows of nonholono\-mic solitonic
pp--waves}

We begin with the ansatz (\ref{5sol2b}) defining a vacuum off--diagonal
solution. That metric does not depend on variable $v$ and possess a Killing
vector $\partial /\partial v.$ It is possible to apply the parametric
transform writing the new family of metrics in terms of polarization
functions,
\begin{eqnarray}
\delta s_{[sol2\theta ^{\prime }]}^{2} &=&-\eta _{2}(\theta ^{\prime })\
dx^{2}+\eta _{3}(\theta ^{\prime })\ dy^{2}-\eta _{4}(\theta ^{\prime })\
h_{0}^{2}\breve{b}^{2}[(qk)^{\ast }]^{2}\delta p^{2}  \notag \\
&&+\eta _{5}(\theta ^{\prime })\ \breve{b}^{2}(qk)^{2}\delta v^{2},  \notag
\\
\delta p &=&dp+\eta _{2}^{4}(\theta ^{\prime })\ \left[ (\ln |qk|)^{\ast }%
\right] ^{-1}\partial _{x}\ln |\breve{b}|\ dx  \notag \\
&&+\eta _{3}^{4}(\theta ^{\prime })\ \left[ (\ln |qk|)^{\ast }\right]
^{-1}\partial _{y}\ln |\breve{b}|\ dy,\   \notag \\
\delta v &=&dv+\eta _{2}^{5}(\theta ^{\prime })n_{2}^{[0]}dx+\eta
_{3}^{5}(\theta ^{\prime })n_{3}^{[0]}dy,  \label{5sol2bgt}
\end{eqnarray}%
where all polarization functions $\eta _{\widehat{\alpha }}(x,y,p,\theta
^{\prime })$ and $\eta _{\widehat{i}}^{a}(x,y,p,\theta ^{\prime })$ depend
on anisotropic coordinate $p,$ labelled by a parameter $\theta ^{\prime }.$
The new class of solutions contains the multiples $q(p)$ and $k(p)$ defined
respectively by solitonic and pp--waves and depends on certain integration
functions like $n_{\widehat{i}}^{[0]}(x,y)$ and integration constant $%
h_{0}^{2}.$ Such values can defined exactly by stating an explicit
coordinate system and for certain boundary and initial conditions.

It should be noted that the metric (\ref{5sol2bgt}) can not be represented
in a form typical for nonholonomic frame vacuum ansatz for the Levi Civita
connection (i.e. it can not be represented, for instance, in the form (78)
with the coefficients satisfying the conditions (79) in Ref. \cite{5vnhes}).
This is obvious because in our case $\eta _{2}$ and $\eta _{3}$ may depend
on ansiotropic coordinates $p,$ i.e. our ansatz is not similar to (\ref%
{5ans5dr}), which is necessary for the anholonomic frame method.
Nevertheless, such classes of metrics define exact vacuum solutions as a
consequence of the Geroch method (for nonholonomic manifolds, we call it
also the method of parametric transforms). This is the priority to consider
together both methods: we can parametrize different type of transforms by
polarization functions in a unified form and in different cases such
polarizations will be subjected to corresponding type of constraints,
generating anholonomic deformations or parametric transforms.

Nevertheless, we can generate nonholonomic Ricci flows solutions from the
very beginning, considering such flows, at the first step of transforms, for
the metric (\ref{5sol2b}), prescribing a constant $\lambda $ necessary for
normalization, which result in (\ref{5sol2bf}) and at the second step to
apply the parametric transforms. After first step, we get an ansatz
\begin{eqnarray}
\delta s_{[sol2a,\chi ]}^{2} &=&-\left[ \eta _{2}(\chi )dx^{2}+\eta
_{3}(\chi )dy^{2}\right]  \notag \\
&& -h_{0}^{2}\breve{b}^{2}(\chi )[(qk)^{\ast }]^{2}\delta p^{2}+\breve{b}%
^{2}(\chi )(qk)^{2}\delta v^{2},  \notag \\
\delta p &=&dp+\left[ (\ln |qk|)^{\ast }\right] ^{-1}\partial _{x}\ln |%
\breve{b}(\chi )|\ dx  \notag \\
&&+\left[ (\ln |qk|)^{\ast }\right] ^{-1}\partial _{y}\ln |\breve{b}(\chi
)|\ dy,\   \notag \\
\delta v &=&dv+n_{2}^{[0]}(\chi )dx+n_{3}^{[0]}(\chi )dy,  \label{5sol2bgta}
\end{eqnarray}%
where we introduced the parametric dependence on $\chi ,$
\begin{equation*}
b(x,y,p,\chi )=\breve{b}(x,y,\chi )q(p)k(p)
\end{equation*}%
which allows us to use the same formulas (\ref{5aux5aa}) for $w_{3,4}$ not
depending on $\chi .$ The values $\breve{b}^{2}(\chi )$ and $%
n_{2}^{[0]}(\chi )$ are constrained to be solutions of
\begin{equation*}
\frac{\partial }{\partial \chi }\left[ -\eta _{2,3}(x,y,\chi )+\breve{b}%
^{2}(x,y,\chi )(q(p)k(p))^{2}~(n_{2,3}^{[0]}(x,y,\chi ))^{2}\right] =0,
\end{equation*}%
obtained by introducing (\ref{5sol2bgta} into (\ref{5eq1}). The second step
is to introduce polarizations functions $\eta _{\widehat{\alpha }%
}(x,y,p,\theta ^{\prime })$ and $\eta _{\widehat{i}}^{a}(x,y,p,\theta
^{\prime })$ for a parametric transform, which is possible because we have a
Killing symmetry on $\partial /\partial v$ and (\ref{5sol2bgta}) is an
Einstein metric (we have to suppose that such parametric transforms can be
defined as solutions of the Geroch equations \cite{5vnhes,5geroch1} for the
Einstein spaces, not only for vacuum metrics, at least for small prescribed
cosmological constants). Finally, we get a two parameter metric, on $\theta
^{\prime }$ and $\chi,$
\begin{eqnarray}
\delta s_{[sol2a,\chi ]}^{2} &=&-\left[ \eta _{2}(\theta ^{\prime
})dx^{2}+\eta _{3}(\theta ^{\prime })dy^{2}\right] -\eta _{4}(\theta
^{\prime })h_{0}^{2}\breve{b}^{2}(\chi )[(qk)^{\ast }]^{2}\delta p^{2}
\label{5sol2bgtaa} \\
&&+\eta _{5}(\theta ^{\prime })\breve{b}^{2}(\chi )(qk)^{2}\delta v^{2},
\notag \\
\delta p &=&dp+\eta _{2}^{4}(\theta ^{\prime })\left[ (\ln |qk|)^{\ast }%
\right] ^{-1}\partial _{x}\ln |\breve{b}|\ dx  \notag \\
&&+\eta _{3}^{4}(\theta ^{\prime })\ \left[ (\ln |qk|)^{\ast }\right]
^{-1}\partial _{y}\ln |\breve{b}|\ dy,\   \notag \\
\delta v &=&dv+\eta _{2}^{5}(\theta ^{\prime })n_{2}^{[0]}(\chi )dx+\eta
_{3}^{5}(\theta ^{\prime })n_{3}^{[0]}(\chi )dy,  \notag
\end{eqnarray}%
with the nonholonomic Ricci flow evolution equations%
\begin{eqnarray*}
&&\frac{\partial }{\partial \chi }\{-\eta _{2,3}(x,y,p,\chi ,\theta ^{\prime
}) + \\
&&\eta _{5}(x,y,p,\theta ^{\prime })\breve{b}^{2}(x,y,\chi )(q(p)k(p))^{2}~%
\left[ \eta _{3}^{5}(x,y,p,\theta ^{\prime })n_{2,3}^{[0]}(x,y,\chi )\right]
^{2}\} =0,
\end{eqnarray*}%
to which one reduces the equation (\ref{5eq1}) by introducing (\ref%
{5sol2bgta}). Such parametric nonholonomic Ricci flows can be constrained
for the Levi Civita connection if we consider coefficients satisfying
certain conditions equivalent to (\ref{5ep2b}) and (\ref{5ep2c}), imposed
for the coefficients of auxiliary metric (\ref{5sol2bgta}), when
\begin{eqnarray*}
\eta _{2}(x,y,\chi ) &=&g_{3}(x,y,\chi )=-e^{\psi (x,y,\chi )}, \\
h_{4} &=&-h_{0}^{2}\breve{b}^{2}(\chi )[(qk)^{\ast }]^{2},h_{5}=\breve{b}%
^{2}(\chi )(qk)^{2}, \\
w_{\widehat{i}} &=&\partial _{\widehat{i}}\phi /\phi ^{\ast },%
\mbox{\ where
\ }\varphi =-\ln \left| \sqrt{|h_{4}h_{5}|}/|h_{5}^{\ast }|\right| , \\
~n_{2,3}(\chi ) &=&n_{2,3}^{[0]}(x,y,\chi )
\end{eqnarray*}%
are constrained to
\begin{eqnarray*}
\psi ^{\bullet \bullet }(\chi )+\psi ^{^{\prime \prime }}(\chi ) &=&-\lambda
\\
h_{5}^{\ast }\phi /h_{4}h_{5} &=&\lambda , \\
w_{2}^{\prime }-w_{3}^{\bullet }+w_{3}w_{2}^{\ast }-w_{2}w_{3}^{\ast } &=&0,
\\
n_{2}^{\prime }(\chi )-n_{3}^{\bullet }(\chi ) &=&0.
\end{eqnarray*}%
In a more general case, we can model Killing---Ricci flows for the canonical
d--connections.

\subsection{Two parameter nonholonomic solitonic pp--waves and flows}

Finally, we give an explicit example of solutions with two parameter $%
(\theta ^{\prime },\theta )$--metrics (see definition of such frame
transform by formulas (81) in Ref. \cite{5vnhes} and (69) in Ref. \cite%
{5nhrf04}). We begin with the ansatz metric $\ _{\shortmid }^{\circ }\mathbf{%
\tilde{g}}_{[2a]}(\theta )$ (\ref{5auxpwc}) with the coefficients subjected
to constraints (\ref{5ep2b}) for $\lambda \rightarrow 0$ and coordinates
parametrized $x^{2}=x,x^{3}=y,y^{4}=p$ and $y^{5}=v.$ We consider that the
solitonic wave $\phi $ is included as a multiple in $\eta _{5}$ and that $%
\kappa =\breve{\kappa}(x,y)k(p)$ is a pp--wave. This family of vacuum
metrics $\ _{\shortmid }^{\circ }\mathbf{\tilde{g}}_{[2a]}\mathbf{(}\theta
\mathbf{)}$ does not depend on variable $v,$ i.e. it possess a Killing
vector $\partial /\partial v,$ which allows us to apply a parametric
transform as we described in the previous example. The resulting two
parameter family of solutions, with redefined polarization functions, is
given by the ansatz
\begin{eqnarray}
\delta s_{[2a]}^{2} &=&-e^{\psi (x,y,\theta )}\left( \overline{\eta }%
_{2}(x,y,p,\theta ^{\prime })dx^{2}+\overline{\eta }_{3}(x,y,p,\theta
^{\prime })dy^{2}\right)  \label{5sol2p2} \\
&&-2\breve{\kappa}(x,y)k(p)\ \eta _{4}(x,y,p,\theta )\overline{\eta }%
_{4}(x,y,p,\theta ^{\prime })\delta p^{2}  \notag \\
&&+\ \frac{\eta _{5}(x,y,p,\theta )\overline{\eta }_{5}(x,y,p,\theta
^{\prime })}{8\breve{\kappa}(x,y)k(p)}\delta v^{2}  \notag \\
\delta p &=&dp+w_{2}(x,y,p,\theta )\overline{\eta }_{2}^{4}(x,y,p,\theta
^{\prime })dx+w_{3}(x,y,p,\theta )\overline{\eta }_{3}^{4}(x,y,p,\theta
^{\prime })dy,\   \notag \\
\delta v &=&dv+n_{2}(x,y,\theta )\overline{\eta }_{2}^{5}(x,y,p,\theta
^{\prime })dx+n_{3}(x,y,\theta )\overline{\eta }_{3}^{5}(x,y,p,\theta
^{\prime })dy.  \notag
\end{eqnarray}%
The set of multiples in the coefficients are parametrized following the
conditions: The value $\breve{\kappa}(x,y)$ is just that defining an exact
vacuum solution for the primary metric (\ref{5auxpw}) stating the first type
of parametric transforms. Then we consider the pp--wave component $k(p)$ and
the solitonic wave included in $\eta _{5}(x,y,p,\theta )$ such way that the
functions $\psi ,\eta _{4,6},w_{2,3}$ and $n_{2,3}$ are subjected to the
condition to define the class of metrics (\ref{5auxpwc}). The metrics are
parametrized both by $\theta ,$ following solutions of the Geroch equations
(see, for instance, the Killing (8) and (9) in Ref. \cite{5nhrf04}), and by
a N--connection splitting with $w_{2,3}$ and $n_{2,3},$ all adapted to the
corresponding nonholonomic deformation derived for $g_{2}(\theta
)=g_{3}(\theta )=e^{\psi (\theta )}$ and $g_{4}=2\breve{\kappa}k\ \eta _{4}$
and $g_{5}=\eta _{5}/8\breve{\kappa}k$ subjected to the conditions (\ref%
{5ep2b})). This set of functions also defines a new set of Killing
equations, for any metric (\ref{5auxpwc}) allowing to find the ''overlined''
polarizations $\overline{\eta }_{\widehat{i}}(\theta ^{\prime })$ and $%
\overline{\eta }_{\widehat{i}}^{a}(\theta ^{\prime }).$

For compatible nonholonomic Ricci flows of the h--metric and
N--connec\-ti\-on coefficients, the class of two parametric vacuum solutions
can be extended for a prescribed value $\lambda $ and new parameter $\chi ,$
\begin{eqnarray}
\delta s_{[2a,\chi ]}^{2} &=&-e^{\psi (x,y,\theta ,\chi )}\left( \overline{%
\eta }_{2}(x,y,p,\theta ^{\prime },\chi )dx^{2}+\overline{\eta }%
_{3}(x,y,p,\theta ^{\prime },\chi )dy^{2}\right)  \notag \\
&&-2\breve{\kappa}(x,y)k(p)\eta _{4}(x,y,p,\theta )\overline{\eta }%
_{4}(x,y,p,\theta ^{\prime })\delta p^{2}  \notag \\
&&+\ \frac{\eta _{5}(x,y,p,\theta )\overline{\eta }_{5}(x,y,p,\theta
^{\prime })}{8\breve{\kappa}(x,y)k(p)}\delta v^{2}  \label{rflsol1} \\
\delta p &=&dp+w_{2}(x,y,p,\theta )\overline{\eta }_{2}^{4}(x,y,p,\theta
^{\prime })dx  \notag \\
&&+w_{3}(x,y,p,\theta )\overline{\eta }_{3}^{4}(x,y,p,\theta ^{\prime })dy,\
\notag \\
\delta v &=&dv+n_{2}(x,y,\theta ,\chi )\overline{\eta }_{2}^{5}(x,y,p,\theta
^{\prime },\chi )dx  \notag \\
&&+n_{3}(x,y,\theta ,\chi )\overline{\eta }_{3}^{5}(x,y,p,\theta ^{\prime
},\chi )dy,  \notag
\end{eqnarray}%
where, for simplicity, we redefine the coefficients in the form
\begin{eqnarray*}
\widetilde{\eta }_{2}(x,y,p,\theta ^{\prime },\theta ,\chi ) &=&e^{\psi
(x,y,\theta ,\chi )}\overline{\eta }_{2}(x,y,p,\theta ^{\prime },\chi )= \\
~\widetilde{\eta }_{3}(x,y,p,\theta ^{\prime },\theta ,\chi ) &=&e^{\psi
(x,y,\theta ,\chi )}\overline{\eta }_{3}(x,y,p,\theta ^{\prime },\chi ),
\end{eqnarray*}
\begin{eqnarray*}
\widetilde{\eta }_{4}(x,y,p,\theta ^{\prime },\theta ) &=&\eta
_{4}(x,y,p,\theta )\overline{\eta }_{4}(x,y,p,\theta ^{\prime }), \\
~\widetilde{\eta }_{5}(x,y,p,\theta ^{\prime },\theta ) &=&\eta
_{5}(x,y,p,\theta )\overline{\eta }_{5}(x,y,p,\theta ^{\prime }), \\
\widetilde{w}_{2,3}(x,y,p,\theta ^{\prime },\theta ) &=&w_{2,3}(x,y,p,\theta
)\overline{\eta }_{2,3}^{4}(x,y,p,\theta ^{\prime }), \\
\widetilde{n}_{2,3}(x,y,p,\theta ^{\prime },\theta ,\chi )
&=&n_{2,3}(x,y,\theta ,\chi )\overline{\eta }_{2,3}^{5}(x,y,p,\theta
^{\prime },\chi ).
\end{eqnarray*}%
Such polarization functions, in general, parametrized by $(\theta ^{\prime
},\theta ,\chi ),$ allow us to write the conditions \ (\ref{5eq1}) for the
classes of metrics (\ref{rflsol1}) in a compact form,%
\begin{equation*}
\frac{\partial }{\partial \chi }\{-\widetilde{\eta }_{2,3}(\theta ^{\prime
},\theta ,\chi )+\frac{\widetilde{\eta }_{5}(\theta ^{\prime },\theta )}{8%
\breve{\kappa}k}~\left[ \widetilde{n}_{2,3}(\theta ^{\prime },\theta ,\chi )%
\right] ^{2}\}=0,
\end{equation*}%
where, for simplicity, we emphasized only the parametric dependencies.

For the configurations with Levi Civita connections, we have to consider
additional constraints of type (\ref{5ep2b}),
\begin{eqnarray*}
\widetilde{\eta }_{2}^{\bullet \bullet }(\chi )+\widetilde{\eta }%
_{2}^{^{\prime \prime }}(\chi ) &=&-\lambda \\
h_{5}^{\ast }\phi /h_{4}h_{5} &=&\lambda , \\
\widetilde{w}_{2}^{\prime }-\widetilde{w}_{3}^{\bullet }+\widetilde{w}_{3}%
\widetilde{w}_{2}^{\ast }-\widetilde{w}_{2}\widetilde{w}_{3}^{\ast } &=&0, \\
n_{2}^{\prime }(x,y,\theta ^{\prime },\theta ,\chi )-n_{3}^{\bullet
}(x,y,\theta ^{\prime },\theta ,\chi ) &=&0
\end{eqnarray*}%
for
\begin{eqnarray*}
\widetilde{h}_{4} &=&-2\breve{\kappa}(x,y)k(p)\widetilde{\eta }%
_{4}(x,y,p,\theta ^{\prime },\theta ),\widetilde{h}_{5}=\widetilde{\eta }%
_{5}(x,y,p,\theta ^{\prime },\theta )/8\breve{\kappa}(x,y)k(p), \\
\widetilde{w}_{\widehat{i}} &=&\partial _{\widehat{i}}\widetilde{\varphi }/%
\widetilde{\varphi }^{\ast },\mbox{\ where \ }\widetilde{\varphi }=-\ln
\left| \sqrt{|h_{4}h_{5}|}/|h_{5}^{\ast }|\right| .
\end{eqnarray*}

The classes of vacuum Einstein metrics (\ref{5sol2p2}) and their, normalized
by a prescribed $\lambda ,$ nonholonomic Ricci flows (\ref{rflsol1}), depend
on certain classes of general functions (nonholonomic and parametric
transform polarizations and integration functions). It is obvious that they
define two parameter $(\theta ^{\prime },\theta)$ nonlinear superpositions
of solitonic waves and pp--waves evolving on parameter $\chi .$ From formal
point of view, the procedure can be iterated for any finite or infinite
number of $\theta $--parameters not depending on coordinates (in principle,
such parameters can depend on flow parameter, but we omit such constructions
in this work). We can construct an infinite number of nonholonomic vacuum
states in gravity, and their possible Ricci flows, constructed from
off--diagonal superpositions of nonlinear waves. Such two transforms do not
commute and depend on order of successive applications.

The nonholonomic deformations not only mix and relate nonlinearly two
different ''Killing'' classes of solutions but introduce into the formalism
the possibility to consider flow evolution configurations and other new very
important and crucial properties. For instance, the polarization functions
can be chosen such ways that the vacuum solutions will posses noncommutative
and/algebroid symmetries even at classical level, or generalized
configurations in order to contain contributions of torsion, nonmetricity
and/or string fields in various generalized models of Lagrange--Hamilton
algebroids, string, brane, gauge, metric--affine and Finsler--Lagrange
gravity, see Refs. \cite{5vesnc,5valg,5vsgg,5vlha}.

\section{Ricci Flows and Parametric Nonholonomic Deformations of the
Schwarzschild Metric}

We construct new classes of exact solutions for Ricci flows and nonholonomic
deformations of the Schwarzschild metric. There are analyzed physical
effects of parametrized families of generic off--diagonal flows and
interactions with solitonic pp--waves.

\subsection{Deformations and flows of stationary backgrounds}

Following the methods outlined Refs. \cite{5nhrf04,5vnhes}, we
nonholonomically deform on angular variable $\varphi $ the Schwarzschild
type solution (\ref{5aux1}) into a generic off--diagonal stationary metric.
For nonholonomic Einstein spaces, we shall use an ansatz  of type
\begin{eqnarray}
\delta s_{[1]}^{2} &=&\epsilon _{1}d\varkappa ^{2}-\eta _{2}(\xi )d\xi
^{2}-\eta _{3}(\xi )r^{2}(\xi )\ d\vartheta ^{2}  \label{5sol1} \\
&&-\eta _{4}(\xi ,\vartheta ,\varphi )r^{2}(\xi )\sin ^{2}\vartheta \ \delta
\varphi ^{2}+\eta _{5}(\xi ,\vartheta ,\varphi )\varpi ^{2}(\xi )\ \delta
t^{2},  \notag \\
\delta \varphi &=&d\varphi +w_{2}(\xi ,\vartheta ,\varphi )d\xi +w_{3}(\xi
,\vartheta ,\varphi )d\vartheta ,\   \notag \\
\delta t &=&dt+n_{2}(\xi ,\vartheta )d\xi +n_{3}(\xi ,\vartheta )d\vartheta ,
\notag
\end{eqnarray}%
where we shall use 3D spacial spherical coordinates, $(\xi (r),\vartheta
,\varphi )$ or $(r,\vartheta ,\varphi ).$ The nonholonomic transform
generating this off--diagonal metric are defined by $g_{i}=\eta _{i}\check{g}%
_{i}$ and $h_{a}=\eta _{a}\check{h}_{a}$ where $(\check{g}_{i},\check{h}%
_{a}) $ are given by data (\ref{5aux1p}).

\subsubsection{General nonholonomic polarizations}

We can construct a class of metrics of type (\ref{5ans5dr}) with the
coefficients subjected to the conditions (\ref{5ep2b}) (in this case, for
the ansatz (\ref{5sol1}) with coordinates $x^{2}=\xi ,x^{3}=\vartheta
,y^{4}=\varphi ,y^{5}=t).$ The solution of (\ref{5ep2a}), for $\lambda =0,$
in terms of polarization functions, can be written
\begin{equation}
\sqrt{|\eta _{4}|}=h_{0}\sqrt{|\frac{\check{h}_{5}}{\check{h}_{4}}|}\left(
\sqrt{|\eta _{5}|}\right) ^{\ast },  \label{5eq23a}
\end{equation}%
where $\check{h}_{a}$ are coefficients stated by the Schwarzschild solution
for the chosen system of coordinates but $\eta _{5}$ can be any function
satisfying the condition $\eta _{5}^{\ast }\neq 0.$ We shall use certain
parametrizations of solutions when%
\begin{eqnarray*}
-h_{0}^{2}(b^{\ast })^{2} &=&\eta _{4}(\xi ,\vartheta ,\varphi )r^{2}(\xi
)\sin ^{2}\vartheta \\
b^{2} &=&\eta _{5}(\xi ,\vartheta ,\varphi )\varpi ^{2}(\xi )
\end{eqnarray*}

The polarizations $\eta _{2}$ and $\eta _{3}$ can be taken in a form that $%
\eta _{2}=\eta _{3}r^{2}=e^{\psi (\xi ,\vartheta ,\chi )},$
\begin{equation*}
\psi ^{\bullet \bullet }+\psi ^{\prime \prime }=0,
\end{equation*}%
defining solutions of (\ref{5ep1a}) for $\lambda =0.$ The solutions of (\ref%
{5ep3a}) and (\ref{5ep4a}) for vacuum configurations of the Levi Civita
connection are constructed
\begin{equation*}
w_{2}=\partial _{\xi }(\sqrt{|\eta _{5}|}\varpi )/\left( \sqrt{|\eta _{5}|}%
\right) ^{\ast }\varpi ,\ w_{3}=\partial _{\vartheta }(\sqrt{|\eta _{5}|}%
)/\left( \sqrt{|\eta _{5}|}\right) ^{\ast }
\end{equation*}%
and any $n_{2,3}(\xi ,\vartheta )$ for which $n_{2}^{\prime }(\chi
)-n_{3}^{\bullet }(\chi )=0.$ For any function $\eta _{5}\sim a_{1}(\xi
,\vartheta )a_{2}(\varphi ),$ the integrability conditions (\ref{5ep2b}) and
(\ref{5ep2c}).

We conclude that the stationary nonholonomic deformations of the
Sch\-warz\-schild metric are defined by the off--diagonal ansatz
\begin{eqnarray}
\delta s_{[1]}^{2} &=&\epsilon _{1}d\chi ^{2}-e^{\psi }\left( d\xi ^{2}+\
d\vartheta ^{2}\right)  \label{5sol1a} \\
&&-h_{0}^{2}\varpi ^{2}\left[ \left( \sqrt{|\eta _{5}|}\right) ^{\ast }%
\right] ^{2}\ \delta \varphi ^{2}+\eta _{5}\varpi ^{2}\ \delta t^{2},  \notag
\\
\delta \varphi &=&d\varphi +\frac{\partial _{\xi }(\sqrt{|\eta _{5}|}\varpi )%
}{\left( \sqrt{|\eta _{5}|}\right) ^{\ast }\varpi }d\xi +\frac{\partial
_{\vartheta }(\sqrt{|\eta _{5}|})}{\left( \sqrt{|\eta _{5}|}\right) ^{\ast }}%
d\vartheta ,\   \notag \\
\delta t &=&dt+n_{2}d\xi +n_{3}d\vartheta ,  \notag
\end{eqnarray}%
where the coefficients do not depend on Ricci flow parameter $\lambda .$
Such vacuum solutions were constructed mapping a static black hole solution
into Einstein spaces with locally anistoropic backgrounds (on coordinate $%
\varphi )$ defined by an arbitrary function $\eta _{5}(\xi ,\vartheta
,\varphi )$ with $\partial _{\varphi }\eta _{5}\neq 0$, an arbitrary $\psi
(\xi ,\vartheta )$ solving the 2D Laplace equation and certain integration
functions $n_{2,3}(\xi ,\vartheta )$ and integration constant $h_{0}^{2}.$
In general, the solutions from the target set of metrics do not define black
holes and do not describe obvious physical situations. Nevertheless, they
preserve the singular character of the coefficient $\varpi ^{2}$ vanishing
on the horizon of a Schwarzschild black hole. We can also consider a
prescribed physical situation when, for instance, $\eta _{5}$ mimics 3D, or
2D, solitonic polarizations on coordinates $\xi ,\vartheta ,\varphi ,$ or on
$\xi ,\varphi .$

For a family of metrics (\ref{5sol1a}), we can consider the ''nearest''
extension to flows of N--connection coefficients $w_{2,3}\rightarrow
w_{2,3}(\chi )$ and $n_{2,3}\rightarrow n_{2,3}(\chi ),$ when for $\lambda
=0,$ and $R_{\alpha \beta }=0,$ the equation (\ref{5eq1}) is satisfied if
\begin{equation}
h_{0}^{2}\left[ \left( \sqrt{|\eta _{5}|}\right) ^{\ast }\right] ^{2}\frac{%
\partial \left( w_{2,3}\right) ^{2}}{\partial \chi }=\eta _{5}\frac{\partial
\left( n_{2,3}\right) ^{2}}{\partial \chi }.  \label{5aux5e}
\end{equation}%
The metric coefficients for such Ricci flows are the same as for the exact
vacuum nonholonomic deformation but with respect to evolving N--adapted dual
basis
\begin{eqnarray}
\delta \varphi (\chi ) &=&d\varphi +w_{2}(\xi ,\vartheta ,\varphi ,\chi
)d\xi +w_{3}(\xi ,\vartheta ,\varphi ,\chi )d\vartheta ,  \label{5fncel} \\
\delta t &=&dt+n_{2}(\xi ,\varphi ,\chi )d\xi +n_{3}(\xi ,\vartheta ,\chi
)d\vartheta ,  \notag
\end{eqnarray}%
with the coefficients being defined by any solution of (\ref{5aux5e}) and (%
\ref{5ep2b}) and (\ref{5ep2c}) for $\lambda =0.$

\subsubsection{Solutions with small nonholonomic polarizations}

\label{5sspe}In a more special case, in order to select physically valuable
configurations, it is better to consider decompositions on a small parameter
$0<\varepsilon <1$ in (\ref{5sol1a}), when
\begin{eqnarray}
\sqrt{|\eta _{4}|} &=&q_{4}^{\hat{0}}(\xi ,\varphi ,\vartheta )+\varepsilon
q_{4}^{\hat{1}}(\xi ,\varphi ,\vartheta )+\varepsilon ^{2}q_{4}^{\hat{2}%
}(\xi ,\varphi ,\vartheta )...,  \notag \\
\sqrt{|\eta _{5}|} &=&1+\varepsilon q_{5}^{\hat{1}}(\xi ,\varphi ,\vartheta
)+\varepsilon ^{2}q_{5}^{\hat{2}}(\xi ,\varphi ,\vartheta )...,  \notag
\end{eqnarray}%
where the ''hat'' indices label the coefficients multiplied to $\varepsilon
,\varepsilon ^{2},...$\footnote{%
Of course, this way we construct not an exact solution, but extract from a
class of exact ones (with less clear physical meaning) certain subclasses of
solutions decomposed (deformed) on a small parameter being related to the
Schwarzschild metric.} The conditions (\ref{5eq23a}) are expressed in the
form
\begin{equation*}
\varepsilon h_{0}\sqrt{|\frac{\check{h}_{5}}{\check{h}_{4}}|\ }\left( q_{5}^{%
\hat{1}}\right) ^{\ast }=q_{4}^{\hat{0}},\ \varepsilon ^{2}h_{0}\sqrt{|\frac{%
\check{h}_{5}}{\check{h}_{4}}|\ }\left( q_{5}^{\hat{2}}\right) ^{\ast
}=\varepsilon q_{4}^{\hat{1}},...
\end{equation*}%
This system can be solved in a form compatible with small decompositions if
we take the integration constant, for instance, to satisfy the condition $%
\varepsilon h_{0}=1$ (choosing a corresponding system of coordinates). For
this class of small deformations, we can prescribe a function $q_{4}^{\hat{0}%
}$ and define $q_{5}^{\hat{1}},$ integrating on $\varphi $ (or inversely,
prescribing $q_{5}^{\hat{1}},$ then taking the partial derivative $\partial
_{\varphi },$ to compute $q_{4}^{\hat{0}}).$ In a similar form, there are
related the coefficients $q_{4}^{\hat{1}}$ and $q_{5}^{\hat{2}}.$ A very
important physical situation is to select the conditions when such small
nonholonomic deformations define rotoid configurations. This is possible,
for instance, if
\begin{equation}
2q_{5}^{\hat{1}}=\frac{q_{0}(r)}{4\mu ^{2}}\sin (\omega _{0}\varphi +\varphi
_{0})-\frac{1}{r^{2}},  \label{5aux1sd}
\end{equation}%
where $\omega _{0}$ and $\varphi _{0}$ are constants and the function $%
q_{0}(r)$ has to be defined by fixing certain boundary conditions for
polarizations. In this case, the coefficient before $\delta t^{2}$ is
approximated in the form
\begin{equation*}
\eta _{5}\varpi ^{2}=1-\frac{2\mu }{r}+\varepsilon (\frac{1}{r^{2}} +2q_{5}^{%
\hat{1}}).
\end{equation*}%
This coefficient vanishes and defines a small deformation of the
Schwarz\-schild spherical horizon into a an ellipsoidal one (rotoid
configuration) given by
\begin{equation*}
r_{+}\simeq \frac{2\mu }{1+\varepsilon \frac{q_{0}(r)}{4\mu ^{2}}\sin
(\omega _{0}\varphi +\varphi _{0})}.
\end{equation*}%
Such solutions with ellipsoid symmetry seem to define static black
ellipsoids (they were investigated in details in Refs. \cite{5vbe1,5vbe2}).
The ellipsoid configurations were proven to be stable under perturbations
and transform into the Schwarzschild solution far away from the ellipsoidal
horizon. This class of vacuum metrics violates the conditions of black hole
uniqueness theorems \cite{5heu} because the ''surface'' gravity is not
constant for stationary black ellipsoid deformations. So, we can construct
an infinite number of ellipsoidal locally anisotropic black hole
deformations. Nevertheless, they present physical interest because they
preserve the spherical topology, have the Minkowski asymptotic and the
deformations can be associated to certain classes of geometric spacetime
distorsions related to generic off--diagonal metric terms. Putting $\varphi
_{0}=0,$ in the limit $\omega _{0}\rightarrow 0,$ we get $q_{5}^{\hat{1}%
}\rightarrow 0$ in (\ref{5aux1sd}). This allows to state the limits $q_{4}^{%
\hat{0}}\rightarrow 1$ for $\varepsilon \rightarrow 0$ in order to have a
smooth limit to the Schwarzschild solution for $\varepsilon \rightarrow 0.$
Here, one must be emphasized that to extract the spherical static black hole
solution is possible if we parametrize, for instance,
\begin{equation*}
\delta \varphi =d\varphi +\varepsilon \frac{\partial _{\xi }(\sqrt{|\eta
_{5}|}\varpi )}{\left( \sqrt{|\eta _{5}|}\right) ^{\ast }\varpi }d\xi
+\varepsilon \frac{\partial _{\vartheta }(\sqrt{|\eta _{5}|})}{\left( \sqrt{%
|\eta _{5}|}\right) ^{\ast }}d\vartheta
\end{equation*}%
and
\begin{equation*}
\delta t=dt+\varepsilon n_{2}(\xi ,\vartheta )d\xi +\varepsilon n_{3}(\xi
,\vartheta )d\vartheta .
\end{equation*}

One can be defined certain more special cases when $q_{5}^{\hat{2}}$ and $%
q_{4}^{\hat{1}}$ (as a consequence) are of solitonic locally anisotropic
nature. In result, such solutions will define small stationary deformations
of the Schwarzschild solution embedded into a background polarized by
anisotropic solitonic waves.

For Ricci flows on N--connection coefficients, such stationary rotoid
configurations evolve with respect to small deformations of co--frames (\ref%
{5fncel}), $\delta \varphi (\chi )$ and $\delta t(\chi ),$ with the
coefficients proportional to $\varepsilon .$

\subsubsection{Parametric nonholonomic transforms of the Schwarz\-schild
solution and their flows}

The ansatz (\ref{5sol1a}) does not depend on time variable and posses a
Killing vector $\partial /\partial t.$ We can apply parametric transforms
and generate families of new solutions depending on a parameter $\theta .$
Following the same steps as for generating (\ref{5sol2bgt}), we construct
\begin{eqnarray}
\delta s_{[1]}^{2} &=&-e^{\psi }\left( \widetilde{\eta }_{2}(\theta )d\xi
^{2}+\ \widetilde{\eta }_{3}(\theta )d\vartheta ^{2}\right)  \label{5sol1b}
\\
&&-h_{0}^{2}\varpi ^{2}\left[ \left( \sqrt{|\eta _{5}|}\right) ^{\ast }%
\right] ^{2}\ \widetilde{\eta }_{4}(\theta )\delta \varphi ^{2}+\eta
_{5}\varpi ^{2}\ \widetilde{\eta }_{5}(\theta )\delta t^{2},  \notag \\
\delta \varphi &=&d\varphi +\widetilde{\eta }_{2}^{4}(\theta )\frac{\partial
_{\xi }(\sqrt{|\eta _{5}|}\varpi )}{\left( \sqrt{|\eta _{5}|}\right) ^{\ast
}\varpi }d\xi +\widetilde{\eta }_{3}^{4}(\theta )\frac{\partial _{\vartheta
}(\sqrt{|\eta _{5}|})}{\left( \sqrt{|\eta _{5}|}\right) ^{\ast }}d\vartheta
,\   \notag \\
\delta t &=&dt+\widetilde{\eta }_{2}^{5}(\theta )n_{2}(\xi ,\vartheta )d\xi +%
\widetilde{\eta }_{3}^{5}(\theta )n_{3}(\xi ,\vartheta )d\vartheta ,  \notag
\end{eqnarray}%
where polarizations $\widetilde{\eta }_{\widehat{\alpha }}(\xi ,\vartheta
,\varphi ,\theta )$ and $\widetilde{\eta }_{\widehat{i}}^{a}(\xi ,\vartheta
,\varphi ,\theta )$ are defined by solutions of the Geroch equations for
Killing symmetries of the vacuum metric (\ref{5sol1a}). Even this class of
metrics does not satisfy the vacuum equations for a typical anholonomic
ansatz, they define vacuum exact solutions and we can apply the formalism on
decomposition on a small parameter $\varepsilon $ like we described in
previous section \ref{5sspe} (one generates not exact solutions, but like in
quantum field theory it can be more easy to formulate a physical
interpretation). For instance, we consider a vacuum background consisting
from solitonic wave polarizations, with components mixed by the parametric
transform, and then to compute nonholonomic deformations of a Schwarzschild
black hole self--consistently imbedded in a such nonperturbative background.

Nonholonomic Ricci flows induced by the N--connection coefficients are given
by flow equations of type (\ref{5fncel}), $\delta \varphi (\chi )$ and $%
\delta t(\chi ),$ with the coefficients depending additionally on $\chi ,$
for instance,
\begin{eqnarray}
w_{2,3}(\xi ,\vartheta ,\varphi ,\theta ,\chi ) &=&\widetilde{\eta }%
_{2,3}^{4}(\xi ,\vartheta ,\varphi ,\theta ,\chi )\frac{\partial _{\xi }(%
\sqrt{|\eta _{5}|}\varpi )}{\left( \sqrt{|\eta _{5}|}\right) ^{\ast }\varpi }%
,  \label{5aux6ee} \\
n_{2,3}(\xi ,\vartheta ,\theta ,\chi ) &=&\widetilde{\eta }_{2}^{5}(\xi
,\vartheta ,\theta ,\chi )n_{2}(\xi ,\vartheta ).  \notag
\end{eqnarray}%
Of course, in order to get Ricci flows with the Levi Civita connection, the
coefficients of (\ref{5sol1b}) and evolving N--connection coefficients (\ref%
{5aux6ee}) have to be additionally constrained by conditions of type (\ref%
{5ep2b}) and (\ref{5ep2c}) for $\lambda =0.$

\subsection{Anisotropic polarizations on extra dimension coordinate}

On can be constructed certain classes of exact off--diagonal solutions when
the extra dimension effectively polarizes the metric coefficients and
interaction constants. We take as a primary metric the ansatz (\ref{5aux2})
(see the parametrization for coordinates for that quadratic element, with $%
x^{1}=\varphi ,$ $x^{2}=\check{\vartheta},$ $x^{3}=\check{\xi},$ $%
y^{4}=\varkappa ,$ $y^{5}=t)$ and consider the off--diagonal target metric
\begin{eqnarray}
\delta s_{[5\varkappa ]}^{2} &=&-r_{g}^{2}\ d\varphi ^{2}-r_{g}^{2}\ \eta
_{2}(\xi ,\check{\vartheta})d\check{\vartheta}^{2}+\eta _{3}(\xi ,\check{%
\vartheta})\check{g}_{3}(\check{\vartheta})\ d\check{\xi}^{2}  \notag \\
&&+\epsilon _{4}\ \eta _{4}(\xi ,\check{\vartheta},\varkappa )\delta
\varkappa ^{2}+\eta _{5}(\xi ,\check{\vartheta},\varkappa )\ \check{h}_{5}\
(\xi ,\check{\vartheta})\ \delta t^{2}  \notag \\
\delta \varkappa &=&d\varkappa +w_{2}(\xi ,\check{\vartheta},\varkappa )d\xi
+w_{3}(\xi ,\check{\vartheta},\varkappa )d\check{\vartheta},\   \label{5sol3}
\\
\delta t &=&dt+n_{2}(\xi ,\check{\vartheta},\varkappa )d\xi +n_{3}(\xi ,%
\check{\vartheta},\varkappa )d\check{\vartheta}.  \notag
\end{eqnarray}%
The coefficients of this ansatz,
\begin{eqnarray*}
g_{1} &=&-r_{g}^{2},g_{2}=-r_{g}^{2}\ \eta _{2}(\xi ,\check{\vartheta}%
),g_{3}=\eta _{3}(\xi ,\check{\vartheta})\check{g}_{3}(\check{\vartheta}), \\
h_{4} &=&\epsilon _{1}\ \eta _{4}(\xi ,\check{\vartheta},\varkappa
),h_{5}=\eta _{5}(\xi ,\check{\vartheta},\varkappa )\ \check{h}_{5}\ (\xi ,%
\check{\vartheta})
\end{eqnarray*}%
are subjected to the condition to solve the system of equations (\ref{5ep1a}%
)--(\ref{5ep4a}) with a nontrivial cosmological constant defined, for
instance, from string gravity by a corresponding ansatz for $H$--fields with
$\lambda =-\lambda _{H}^{2}/2,$ or other type cosmological constants, see
details on such nonholonomic configurations in Refs. \cite{5nhrf04,5vnhes}.

The general solution is given by the data
\begin{equation}
-r_{g}^{2}\ \eta _{2}=\eta _{3}\check{g}_{3}=\exp 2\psi (\xi ,\check{%
\vartheta}),  \label{5auxx1}
\end{equation}%
where $\psi $ is the solution of
\begin{equation*}
\psi ^{\bullet \bullet }+\psi ^{\prime \prime }=\lambda ,
\end{equation*}%
\begin{eqnarray}
\ \eta _{4}\ &=&h_{0}^{2}(\xi ,\check{\vartheta})\left[ f^{\ast }(\xi ,%
\check{\vartheta},\chi )\right] ^{2}|\varsigma (\xi ,\check{\vartheta}%
,\varkappa )|\   \label{5auxx2} \\
\eta _{5}\ \check{h}_{5}\ &=&\left[ f(\xi ,\check{\vartheta},\varkappa
)-f_{0}(\xi ,\check{\vartheta})\right] ^{2},  \notag
\end{eqnarray}%
where
\begin{eqnarray*}
\varsigma (\xi ,\check{\vartheta},\chi ) &=&\varsigma _{\lbrack 0]}(\xi ,%
\check{\vartheta}) \\
&&+\frac{\epsilon _{4}}{16}h_{0}^{2}(\xi ,\check{\vartheta})\lambda
_{H}^{2}\int f^{\ast }(\xi ,\check{\vartheta},\chi )\left[ f(\xi ,\check{%
\vartheta},\chi )-f_{0}(\xi ,\check{\vartheta})\right] d\chi .
\end{eqnarray*}%
The N--connection coefficients $N_{i}^{4}=w_{i}(\xi ,\check{\vartheta}%
,\varkappa ),\ N_{i}^{5}=n_{i}(\xi ,\check{\vartheta},\varkappa )$ are
computed following the formulas
\begin{eqnarray}
w_{\widehat{i}}&=&-\frac{\partial _{\widehat{i}}\varsigma (\xi ,\check{%
\vartheta},\varkappa )}{\varsigma ^{\ast }(\xi ,\check{\vartheta},\varkappa )%
}  \label{53auxx3} \\
n_{\widehat{k}}&=&n_{\widehat{k}[1]}(\xi ,\check{\vartheta})+n_{\widehat{k}%
[2]}(\xi ,\check{\vartheta})\int \frac{\left[ f^{\ast }(\xi ,\check{\vartheta%
},\varkappa )\right] ^{2}}{\left[ f(\xi ,\check{\vartheta},\varkappa
)-f_{0}(\xi ,\check{\vartheta})\right] ^{3}}\varsigma (\xi ,\check{\vartheta}%
,\varkappa )d\varkappa .  \label{53auxx4}
\end{eqnarray}

The solutions depend on arbitrary nontrivial functions $f(\xi ,\check{%
\vartheta},\varkappa )$ (with $f^{\ast }\neq 0),$ $f_{0}(\xi ,\check{%
\vartheta}),$ $h_{0}^{2}(\xi ,\check{\vartheta}),$ $\ \varsigma _{\lbrack
0]}(\xi ,\check{\vartheta}),$ $n_{k[1]}(\xi ,\check{\vartheta})$ and $\
n_{k[2]}(\xi ,\check{\vartheta}),$ and value of cosmological constant $%
\lambda .$ These values have to be defined by certain boundary conditions
and physical considerations. In the sourceless case, $\varsigma _{\lbrack
0]}\rightarrow 1.$\ For the Levi Civita connection, we have to consider $%
h_{0}^{2}(\xi ,\check{\vartheta})\rightarrow const$ and have to prescribe
the integration functions of type $n_{\widehat{k}[2]}=0$ and $n_{\widehat{k}%
[1]}$ solving the equation $\partial _{\check{\vartheta}}n_{2[1]}=\partial
_{\xi }n_{3[1]},$ in order to satisfy some conditions of type (\ref{5ep2b})
and (\ref{5ep2c}).

The class of solutions (\ref{5sol3}) define self--consistent nonholonomic
maps of the Schwarzschild solution into a 5D backgrounds with nontrivial
sources, depending on a general function $f(\xi ,\check{\vartheta},\varkappa
)$ and mentioned integration functions and constants. Fixing $f(\xi ,\check{%
\vartheta},\varkappa )$ to be a 3D soliton (we can consider also solitonic
pp--waves as in previous sections) running on extra dimension $\varkappa ,$
we describe self-consisted embedding of the Schwarzschild solutions into
nonlinear wave 5D curved spaces. In general, it is not clear if any target
solutions preserve the black hole character of primary solution. It is
necessary a very rigorous analysis of geodesic configurations on such
spacetimes, definition of horizons, singularities and so on. Nevertheless,
for small nonholonomic deformations (by introducing a small parameter $%
\varepsilon ,$ like in the section \ref{5sspe}), we can select classes of
''slightly'' deformed solutions preserving the primary black hole character.
In 5D, such solutions are not subjected to the conditions of black hole
uniqueness theorems, see \cite{5heu} and references therein.

The ansatz (\ref{5sol3}) posses two Killing vector symmetries, on $\partial
/\partial t$ and $\partial /\partial \varphi .$ In the sourceless case, we
can apply a parametric transform and generate new families depending on a
parameter $\theta ^{\prime }.$ The constructions are similar to those
generating (\ref{5sol1b}) (we omit here such details). Here we emphasize
that we can not apply a parametric transform to the primary metric (\ref%
{5aux2}) (it is not a vacuum solution) in order to generate families of
parametric solutions with the aim to subject them to further anholonomic
transforms.

For nontrivial cosmological constant (normalization), the metric (\ref{5sol3}%
) can be generalized for nonholonomic Ricci flows of type
\begin{eqnarray}
\delta s_{[5\varkappa \chi ]}^{2} &=&-r_{g}^{2}\ d\varphi ^{2}-r_{g}^{2}\
\eta _{2}(\xi ,\check{\vartheta},\chi )d\check{\vartheta}^{2}+\eta _{3}(\xi ,%
\check{\vartheta},\chi )\check{g}_{3}(\check{\vartheta})\ d\check{\xi}^{2}
\notag \\
&&+\epsilon _{4}\ \eta _{4}(\xi ,\check{\vartheta},\varkappa )\delta
\varkappa ^{2}+\eta _{5}(\xi ,\check{\vartheta},\varkappa )\ \check{h}_{5}\
(\xi ,\check{\vartheta})\ \delta t^{2}  \notag \\
\delta \varkappa &=&d\varkappa +w_{2}(\xi ,\check{\vartheta},\varkappa )d\xi
+w_{3}(\xi ,\check{\vartheta},\varkappa )d\check{\vartheta},\
\label{5sol3a} \\
\delta t(\chi ) &=&dt+n_{2}(\xi ,\check{\vartheta},\chi )d\xi +n_{3}(\xi ,%
\check{\vartheta},\varkappa ,\chi )d\check{\vartheta}.  \notag
\end{eqnarray}%
where the equation (\ref{5eq1}) imposes constraints of type (\ref{5eq1b})
\begin{equation*}
\frac{\partial }{\partial \chi }\left[ g_{2,3}(\xi ,\check{\vartheta},\chi
)+h_{5}(\xi ,\check{\vartheta},\varkappa )~(n_{2,3}(\xi ,\check{\vartheta}%
,\varkappa ,\chi ))^{2}\right] =0,
\end{equation*}%
with is very different from constraints of type (\ref{5aux5e}), for
\begin{eqnarray*}
g_{2} &=&-r_{g}^{2}\ \eta _{2}(\xi ,\check{\vartheta},\chi ),g_{3}=\eta
_{3}(\xi ,\check{\vartheta},\chi )\check{g}_{3}(\check{\vartheta}),\
h_{5}=\eta _{5}(\xi ,\check{\vartheta},\varkappa )\ \check{h}_{5}\ (\xi ,%
\check{\vartheta}), \\
n_{\widehat{k}}(\chi ) &=&n_{\widehat{k}}^{[1]}(\xi ,\check{\vartheta},\chi
)+n_{\widehat{k}}^{[2]}(\xi ,\check{\vartheta},\chi)\int \frac{\left[
f^{\ast }(\xi ,\check{\vartheta},\varkappa )\right] ^{2}}{[ f(\xi ,\check{%
\vartheta},\varkappa )-f_{0}(\xi ,\check{\vartheta})] ^{3}}\varsigma (\xi ,%
\check{\vartheta},\varkappa )d\varkappa .
\end{eqnarray*}%
For holonomic Ricci flows with the Levi Civita connection, we have to
consider additional constraints%
\begin{eqnarray}
\psi ^{\bullet \bullet }(\chi )+\psi ^{^{\prime \prime }}(\chi ) &=&\lambda
\label{aux6zz} \\
h_{5}^{\ast }\phi /h_{4}h_{5} &=&\lambda ,  \notag \\
w_{2}^{\prime }-w_{3}^{\bullet }+w_{3}w_{2}^{\ast }-w_{2}w_{3}^{\ast } &=&0,
\notag \\
n_{2}^{\prime }(\chi )-n_{3}^{\bullet }(\chi ) &=&0.  \notag
\end{eqnarray}%
for
\begin{eqnarray*}
g_{2}(\xi ,\check{\vartheta},\chi ) &=&g_{3}(\xi ,\check{\vartheta},\chi
)=e^{2\psi (\xi ,\check{\vartheta},\chi )},h_{4}=\epsilon _{4}\ \eta
_{4}(\xi ,\check{\vartheta},\varkappa ), \\
w_{\widehat{i}} &=&\partial _{\widehat{i}}\phi /\phi ^{\ast },%
\mbox{\ where
\ }\varphi =-\ln \left| \sqrt{|h_{4}h_{5}|}/|h_{5}^{\ast }|\right| , \\
~n_{2,3}(\chi ) &=&n_{2,3}^{[1]}(\xi ,\check{\vartheta},\chi ).
\end{eqnarray*}%
This class of Ricci flows, defined by the family of solutions (\ref{5sol3a})
describes deformed Schwarzschild metrics, running on extra dimension
coordinate $\varkappa $ with mutually compatible evolution of the
h--component of metric and the $n$--coefficients of the N--connection.

\subsection{5D solutions with nonholonomic time like coordinate}

We use the primary metric (\ref{5aux3}) (which is not a vacuum solution and
does not admit parametric transforms but can be nonholonomically deformed)
resulting in a target off--diagonal ansatz,
\begin{eqnarray}
\delta s_{[3t]}^{2} &=&-r_{g}^{2}\ d\varphi ^{2}-r_{g}^{2}\eta _{2}(\xi ,%
\check{\vartheta})\ d\check{\vartheta}^{2}+\eta _{3}(\xi ,\check{\vartheta})%
\check{g}_{3}(\check{\vartheta})\ d\check{\xi}^{2}  \notag \\
&&+\eta _{4}(\xi ,\check{\vartheta},t)\ \check{h}_{4}\ (\xi ,\check{\vartheta%
})\ \delta t^{2}+\epsilon _{5}\ \eta _{5}(\xi ,\check{\vartheta},t)\ \delta
\varkappa ^{2},  \notag \\
\delta t &=&dt+w_{2}(\xi ,\check{\vartheta},t)d\xi +w_{3}(\xi ,\check{%
\vartheta},t)d\check{\vartheta},  \label{5sol4} \\
\delta \varkappa &=&d\varkappa +n_{2}(\xi ,\check{\vartheta},t)d\xi
+n_{3}(\xi ,\check{\vartheta},t)d\check{\vartheta},\   \notag
\end{eqnarray}%
where the local coordinates are established $x^{1}=\varphi ,\ x^{2}=\check{%
\vartheta},\ x^{3}=\check{\xi},\ y^{4}=t,\ y^{5}=\varkappa $ and the
polarization functions and coefficients of the N--connection are chosen to
solve the system of equations (\ref{5ep1a})--(\ref{5ep4a}). Such solutions
are generic 5D and emphasize the anisotropic dependence on time like
coordinate $t.$ The coefficients can be computed by the same formulas (\ref%
{5auxx1}) and (\ref{5auxx2}) as in the previous section, for the ansatz (\ref%
{5sol3}), by changing the coordinate $t$ into $\varkappa $ and, inversely, $%
\varkappa $ into $t.$ This class of solutions depends on a function $f(\xi ,%
\check{\vartheta},t),$ with $\partial _{t}f\neq 0,$ and on integration
functions (depending on $\xi $ and $\check{\vartheta}$) and constants. We
can consider more particular physical situations when $f(\xi ,\check{%
\vartheta},t)$ defines a 3D solitonic wave, or a pp--wave, or their
superpositions, and analyze configurations when a Schwarzschild black hole
is self--consistently embedded into a dynamical 5D background. We analyzed
certain similar physical situations in Ref. \cite{5vs1} when an extra
dimension soliton ''running'' away a 4D black hole.

The set of 5D solutions (\ref{5sol4}) posses two Killing vector symmetry, $%
\partial /\partial t$ and $\partial /\partial \varkappa ,$ like in the
previous section, but with another types of vectors. For the vacuum
configurations, it is possible to perform a parametric transform and
generate parametric (on $\theta ^{\prime }$) 5D solutions (labelling, for
instance, packages of nonlinear waves).

For nontrivial cosmological constant (normalization), the metric (\ref{5sol4}%
) also can be generalized to describe nonholonomic Ricci flows
\begin{eqnarray}
\delta s_{[3t,\chi ]}^{2} &=&-r_{g}^{2}\ d\varphi ^{2}-r_{g}^{2}\eta
_{2}(\xi ,\check{\vartheta},\chi )\ d\check{\vartheta}^{2}+\eta _{3}(\xi ,%
\check{\vartheta},\chi )\check{g}_{3}(\check{\vartheta})\ d\check{\xi}^{2}
\notag \\
&&+\eta _{4}(\xi ,\check{\vartheta},t)\ \check{h}_{4}\ (\xi ,\check{\vartheta%
})\ \delta t^{2}+\epsilon _{5}\ \eta _{5}(\xi ,\check{\vartheta},t)\ \delta
\varkappa ^{2},  \notag \\
\delta t &=&dt+w_{2}(\xi ,\check{\vartheta},t)d\xi +w_{3}(\xi ,\check{%
\vartheta},t)d\check{\vartheta},  \label{5sol4af} \\
\delta \varkappa (\chi ) &=&d\varkappa +n_{2}(\xi ,\check{\vartheta},t,\chi
)d\xi +n_{3}(\xi ,\check{\vartheta},t,\chi )d\check{\vartheta},\   \notag
\end{eqnarray}%
where the equation (\ref{5eq1}) imposes constraints of type (\ref{5eq1b})
\begin{equation*}
\frac{\partial }{\partial \chi }\left[ g_{2,3}(\xi ,\check{\vartheta},\chi
)+h_{5}(\xi ,\check{\vartheta},\varkappa )~(n_{2,3}(\xi ,\check{\vartheta}%
,\varkappa ,\chi ))^{2}\right] =0,
\end{equation*}
for
\begin{eqnarray*}
g_{2} &=&-r_{g}^{2}\ \eta _{2}(\xi ,\check{\vartheta},\chi ),g_{3}=\eta
_{3}(\xi ,\check{\vartheta},\chi )\check{g}_{3}(\check{\vartheta}),\
h_{5}=\epsilon _{5}\eta _{5}(\xi ,\check{\vartheta},t)\ , \\
n_{\widehat{k}}(\chi ) &=&n_{\widehat{k}}^{[1]}(\xi ,\check{\vartheta},\chi
)+n_{\widehat{k}}^{[2]}(\xi ,\check{\vartheta},,\chi )\int \frac{\left[
f^{\ast }(\xi ,\check{\vartheta},t)\right] ^{2}}{\left[ f(\xi ,\check{%
\vartheta},t)-f_{0}(\xi ,\check{\vartheta})\right] ^{3}}\varsigma (\xi ,%
\check{\vartheta},t)dt.
\end{eqnarray*}%
For holonomic Ricci flows with the Levi Civita connection, we have to
consider additional constraints of type (\ref{aux6zz}) with re--defined
coefficients and coordinates, when
\begin{eqnarray*}
g_{2}(\xi ,\check{\vartheta},\chi ) &=&g_{3}(\xi ,\check{\vartheta},\chi
)=e^{2\psi (\xi ,\check{\vartheta},\chi )},h_{4}=\eta _{4}(\xi ,\check{%
\vartheta},t)\ \check{h}_{4}\ (\xi ,\check{\vartheta}), \\
w_{\widehat{i}} &=&\partial _{\widehat{i}}\phi /\phi ^{\ast },%
\mbox{\ where
\ }\varphi =-\ln \left| \sqrt{|h_{4}h_{5}|}/|h_{5}^{\ast }|\right| , \\
~n_{2,3}(\chi ) &=&n_{2,3}^{[1]}(\xi ,\check{\vartheta},\chi ).
\end{eqnarray*}%
This class of Ricci flows, defined by the family of solutions (\ref{5sol4af}%
) describes deformed Schwarzschild metrics, running on time like coordinate $%
t$ with mutually compatible evolution of the h--component of metric and the $%
n$--coefficients of the N--connection.

\section{Discussion}

We constructed exact solutions in gravity and Ricci flow theory following
superpositions of the parametric and anholonomic frame transforms. A
geometric method previously elaborated in our partner works \cite%
{5nhrf04,5vnhes} was applied to generalizations of valuable physical
solutions (like solitonic waves, pp--waves and Schwarzschild metrics) in
vacuum gravity. In this work, our investigations were restricted to
nonholonomic Ricci flows of the mentioned type solutions modelled with
respect certain classes of compatible metric and associated nonlinear
connection (N--connection) coefficients when the solutions of evolution/
field equations can be obtained in general form.

The first advance is the possibility to generalize vacuum metrics by
allowing realistic string gravity or matter field sources which can be
encoded as an effective (in general, nonhomogeneous) cosmological constant
on nonholonomic (pseudo) Riemannian spaces of dimensions four and five (4D
and 5D) and deriving nonlinear solitonic and pp--wave interactions and their
Ricci flows.

The second kind of progress is the proof of existence of multi--parametric
transforms, associated to certain Killing symmetries, like the Geroch
equations \cite{5geroch1,5geroch2}, mapping certain target metrics (in our
case of physical importance) into different classes of generic off--diagonal
exact solutions admitting different scenarios of Ricci flows depending on
the type of nonholonomic frame constraints.

The outcome of the first advance is rather satisfactory: we can in a similar
way consider parametric deformations of metrics and flows of geometric and
physical objects by obtaining, for instance, static rotoid configurations,
solitonic and pp--wave propagation of black holes on time like and extra
dimension coordinates.

However, the outcome of the second kind of progress raises as many problems
as it solves: we should provide a physical motivation for the
multi--parameter dependence and 'hidden' Killing symmetry under nonholonomic
deformations. If one of the parameters is identified with the Ricci flow
parameter, it may be considered to describe a corresponding evolution. In
general, this may be associated to chains of Ricci multi--flows but not
obligatory. We have to speculate additionally on physical meaning of such
parametric solutions both for vacuum gravitational and Einstein spaces and
in Ricci flow theory when metrics and connections are subjected to
nonholonomic constraints on coefficients and associated frames.

Whereas most previous work on Ricci flow theory and applications has
concentrated on some approximate methods and simplest classes of solutions,
the present paper aims to elaboration of general geometric methods of
constructing solutions and deriving their physically important symmetries.
There were stated exact principles how the physically important solutions in
gravity theories can be deformed in multi--parametric ways to describe
off--diagonal nonlinear gravitational and matter field interactions and the
evolution of physical and geometric objects. The first step was to derive
exact solutions in the most possible general form preserving dependence not
only on transform and flow parameters but on classes of generating and
integration functions and constants. Further work would be needed to analyze
more rigorously certain important physical effects with exactly defined
boundary and initial conditions when the integration functions and constants
are defined in explicit form.

\vskip5pt

\textbf{Acknowledgement:} The work is performed during a visit at Fields
Institute.

\appendix

\setcounter{equation}{0} \renewcommand{\theequation}
{A.\arabic{equation}} \setcounter{subsection}{0}
\renewcommand{\thesubsection}
{A.\arabic{subsection}}

\section{Cosmological Constants and Strings}

The simplest way to perform a local covariant calculus by applying
d--connecti\-ons is to use N--adapted differential forms and to introduce
the d--connection 1--form $\mathbf{\Gamma }_{\ \beta }^{\alpha }=\mathbf{%
\Gamma }_{\ \beta \gamma }^{\alpha }\mathbf{e}^{\gamma },$ when the
N--adapted components of d-connection $\mathbf{D}_{\alpha }=(\mathbf{e}%
_{\alpha }\rfloor \mathbf{D})$ are computed following formulas
\begin{equation}
\mathbf{\Gamma }_{\ \alpha \beta }^{\gamma }\left( u\right) =\left( \mathbf{D%
}_{\alpha }\mathbf{e}_{\beta }\right) \rfloor \mathbf{e}^{\gamma },
\label{5cond2}
\end{equation}%
where ''$\rfloor "$ denotes the interior product. This allows us to define
in N--adapted torsion $\mathbf{T=\{\mathcal{T}^{\alpha }\}}$,
\begin{equation}
\mathcal{T}^{\alpha }\doteqdot \mathbf{De}^{\alpha }=d\mathbf{e}^{\alpha }+%
\mathbf{\Gamma }_{\ \beta }^{\alpha }\wedge \mathbf{e}_{\alpha },
\label{5torsa}
\end{equation}%
and curvature $\mathbf{R}=\{\mathcal{R}_{\ \beta }^{\alpha }\},$
\begin{equation*}
\mathcal{R}_{\ \beta }^{\alpha }\doteqdot \mathbf{D\Gamma }_{\beta }^{\alpha
}=d\mathbf{\Gamma }_{\beta }^{\alpha }-\mathbf{\Gamma }_{\ \beta }^{\gamma
}\wedge \mathbf{\Gamma }_{\ \gamma }^{\alpha }.
\end{equation*}

In string gravity, the nontrivial torsion components and string corrections
to matter sources in the Einstein equations can be related to certain
effective interactions with the strength (torsion)
\begin{equation*}
H_{\mu \nu \rho }=\mathbf{e}_{\mu }B_{\nu \rho }+\mathbf{e}_{\rho }B_{\mu
\nu }+\mathbf{e}_{\nu }B_{\rho \mu }
\end{equation*}%
of an antisymmetric field $B_{\nu \rho },$ when%
\begin{equation}
R_{\mu \nu }=-\frac{1}{4}H_{\mu }^{\ \nu \rho }H_{\nu \lambda \rho }
\label{5c01}
\end{equation}%
and
\begin{equation}
D_{\lambda }H^{\lambda \mu \nu }=0,  \label{5c02}
\end{equation}%
see details on string gravity, for instance, in Refs. \cite%
{5string1,5string2}. The conditions (\ref{5c01}) and (\ref{5c02}) are
satisfied by the ansatz
\begin{equation}
H_{\mu \nu \rho }=\widehat{Z}_{\mu \nu \rho }+\widehat{H}_{\mu \nu \rho
}=\lambda _{\lbrack H]}\sqrt{\mid g_{\alpha \beta }\mid }\varepsilon _{\nu
\lambda \rho }  \label{5ansh}
\end{equation}%
where $\varepsilon _{\nu \lambda \rho }$ is completely antisymmetric and the
distorsion (from the Levi--Civita connection) and
\begin{equation*}
\widehat{Z}_{\mu \alpha \beta }\mathbf{c}^{\mu }=\mathbf{e}_{\beta }\rfloor
\mathcal{T}_{\alpha }-\mathbf{e}_{\alpha }\rfloor \mathcal{T}_{\beta }+\frac{%
1}{2}\left( \mathbf{e}_{\alpha }\rfloor \mathbf{e}_{\beta }\rfloor \mathcal{T%
}_{\gamma }\right) \mathbf{c}^{\gamma }
\end{equation*}%
is defined by the torsion tensor (\ref{5torsa}), which for the canonical
d--connection is induced by the coefficients of N--connection, see details
in \cite{5vesnc,5vsgg,5nhrf01,5nhrf02}. We emphasize that our $H$--field
ansatz is different from those formally used in string gravity when $%
\widehat{H}_{\mu \nu \rho }=\lambda _{\lbrack H]}\sqrt{\mid g_{\alpha \beta
}\mid }\varepsilon _{\nu \lambda \rho }.$ \ In our approach, we define $%
H_{\mu \nu \rho }$ and $\widehat{Z}_{\mu \nu \rho }$ from the respective
ansatz for the $H$--field and nonholonomically deformed metric, compute the
torsion tensor for the canonical distinguished connection and, finally,
define the 'deformed' H--field as $\widehat{H}_{\mu \nu \rho }=\lambda
_{\lbrack H]}\sqrt{\mid g_{\alpha \beta }\mid }\varepsilon _{\nu \lambda
\rho }-\widehat{Z}_{\mu \nu \rho }.$

\end{document}